\documentclass[final,superscriptaddress,english,twocolumn,amssymb,aps,prl,
longbibliography]{revtex4-2}
\usepackage{graphicx}
\usepackage{natbib}
\usepackage{amsmath}
\usepackage{amssymb}
\usepackage{wasysym}
\usepackage{oplotsymbl}
\usepackage{appendix}
\usepackage{soul}
\usepackage[section]{placeins}
\definecolor{darkblue}{rgb}{0,0,.65}
\definecolor{darkgreen}{rgb}{0.28,0.41,0.19}
\definecolor{nicegreen}{rgb}{0.28,0.85,0.19}
\usepackage{lipsum}
\usepackage{nicefrac}
\usepackage[%
  pdfauthor={Robin Schaefer},%
  pdfstartview=FitH,%
  breaklinks=true,%
  bookmarks=true,%
  colorlinks=true,%
  anchorcolor=black,%
  citecolor=darkgreen,
  filecolor=black,%
  menucolor=black,%
  urlcolor=darkblue,%
  linkcolor=blue,%
 ]{hyperref}
\usepackage{diagbox}

\def\equationautorefname~#1\null{Eq. (#1)\null}

\newcommand{\appref}[1]{\hyperref[#1]{App.~\ref*{#1}}}
\usepackage[all]{hypcap} %
\usepackage{physics}

\newcommand{\braOket}[3]{\langle\,#1\, \vert \, #2\,\vert\,#3\rangle}

\renewcommand\vec{\mathbf}

\usepackage[normalem]{ulem}

\newcommand{\cf}{\textit{cf.} }

\graphicspath{{images/}}
\begin{document}

\title{\textcolor{}}

\title{Abundance of Hard-Hexagon Crystals in the Quantum Pyrochlore Antiferromagnet}

\author{Robin Sch\"afer}\email{schaefer@pks.mpg.de}
\affiliation{Max Planck Institute for the Physics of Complex Systems, 
Noethnitzer Str. 38, 01187 Dresden, Germany}
\author{Benedikt Placke}
\affiliation{Max Planck Institute for the Physics of Complex Systems, 
Noethnitzer Str. 38, 01187 Dresden, Germany}
\email{placke@pks.mpg.de}

\author{Owen Benton}\email{benton@pks.mpg.de}
\affiliation{Max Planck Institute for the Physics of Complex Systems, 
Noethnitzer Str. 38, 01187 Dresden, Germany}

\author{Roderich Moessner}\email{moessner@pks.mpg.de}\affiliation{Max Planck Institute for the Physics of Complex Systems, 
Noethnitzer Str. 38, 01187 Dresden, Germany}

\date{\today}

\begin{abstract}
We propose a simple family of valence-bond crystals as potential ground states of the $S=1/2$ and $S=1$ Heisenberg antiferromagnet on the pyrochlore lattice. Exponentially numerous in the linear size of the system, these can be visualized as hard-hexagon coverings, with each hexagon representing a resonating valence-bond ring. This ensemble spontaneously  breaks rotation, inversion and translation symmetries. A simple, yet accurate, variational wave function allows a precise determination of the energy, confirmed by density matrix renormalization group and numerical linked cluster expansion, and extended by an analysis of excited states. The identification of the origin of the stability indicates applicability to a broad class of frustrated lattices,
which we demonstrate for the checkerboard and ruby lattices.
Our work suggests a perspective on such quantum magnets, in which {\it unfrustrated} motifs are effectively uncoupled by the frustration of their interactions.  
\end{abstract} 

\maketitle

The propensity of frustrated quantum magnets to host exotic ground states makes them 
a rewarding target for study. For several of the most prominent models, however,
it has proved difficult to come to a consensus about the exact nature of the ground state
\cite{marston_spinpeierls_1991, singh_1992, sachdev1992, chalker_ground_1992, yang1993,
waldtmann98,bernhard2002, budnik2004, singh_ground_2007, hermele_properties_2008,jiang_density_2008,poilblanc2010, yan_science_2011, depenbrock_prl_2012, iqbal2013, liao2017, he2017, lauchli_sfrac12_2019, lawler_2008_1, lawler_2008_2, bergholtz_2010, chubukov1992, deutscher93, hu2015, zhu2015, iqbal16, canals_pyrochlore_1998, canals_lacroix_prb_2000, kim_prb_2008, burnell_monopole_2009,lee_2012, iqbal_quantum_2019, Tsunetsugu_pyro_2001, isoda_valence_bond_1998, harris_ordering_1991, berg_singlet_2003, 
hagymasi_possible_2021, astrakhantsev_broken-symmety_2021, Hering2022,
benton_nlce_pyrochlore_qsl_2018}.
This is particularly true for frustrated isotropic Heisenberg models with low spin $S\leq1$ in dimension $d> 1$, for
which approximate analytical methods are generally uncontrolled and numerical studies are challenging. 
To improve the situation requires simple yet reliable heuristics, by which we can understand
not just which state is the ground state but also whence it derives its stability.

\begin{figure}
    \centering
    \includegraphics{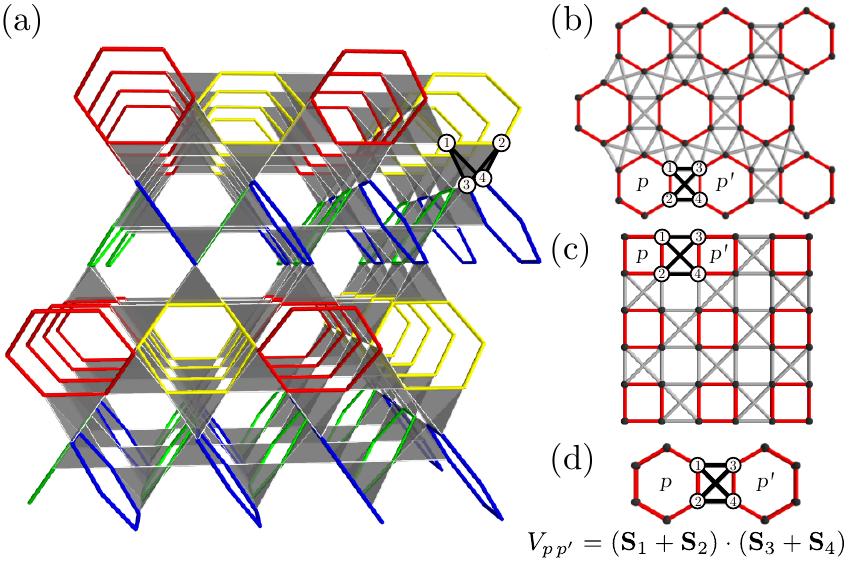}
    \caption{Lattices composed of unfrustrated motifs coupled via a quartet of frustrated bonds. (a) A hard-hexagon tiling \cite{lee_emergent_2002} of the pyrochlore showing $2\times 2\times 2$ unit cells, each containing $48$ sites. {{The different colors represent the four different orientations of hexagons.}} (b) The ruby lattice with additional frustrated couplings exhibiting a unique hard-hexagon tiling. (c) The checkerboard lattice shows one of two possible tilings with unfrustrated squares. {{(d) Illustration of doubly frustrated interactions between motifs: two pairs of antiferromagnetically correlated spins are symmetrically coupled, making the interaction effectively a product of two small terms.}}}
    \label{fig:hard_hexaggon_cover}
\end{figure}

Amongst frustrated lattices, the pyrochlore lattice -- a network of corner-sharing tetrahedra [\autoref{fig:hard_hexaggon_cover}(a)] -- is a particularly tricky case. 
Several proposals have been made for the ground state of the nearest neighbor pyrochlore Heisenberg model
for $S=1/2$ including various forms of quantum spin liquid (QSL) \cite{canals_pyrochlore_1998, canals_lacroix_prb_2000, kim_prb_2008, burnell_monopole_2009,lee_2012, iqbal_quantum_2019}, valence-bond solids \cite{Tsunetsugu_pyro_2001, isoda_valence_bond_1998, harris_ordering_1991, berg_singlet_2003, hagymasi_possible_2021, hagymasi_magnetization_2022, astrakhantsev_broken-symmety_2021, Hering2022, schnack_magnetism_2018}, or the possibility that it lies on a phase boundary between different QSLs \cite{benton_nlce_pyrochlore_qsl_2018}.
Two recent numerical studies found evidence of inversion symmetry breaking \cite{hagymasi_possible_2021, astrakhantsev_broken-symmety_2021},
consistent with one of the earliest proposals \cite{harris_ordering_1991}.

In this Letter, we propose a family of valence-bond crystal states generated from hard (nonoverlapping) hexagon coverings of the pyrochlore lattice \cite{lee_emergent_2002} shown in \autoref{fig:hard_hexaggon_cover}(a). A one-parameter variational wave function describing these states achieves an energy equal to the best-known proposals to date within numerical uncertainties. The hard-hexagon crystal states are exponentially numerous in the linear system size, demonstrating the abundance of competing low-energy states. The 
difficulty in arriving at a consensus over the ground state is likely in considerable part due to the presence of so many competing states
with barely distinguishable energies.

The simplicity of the hard-hexagon states enables us to understand their low energy and stability, which arises from the following ingredients.
The first is that the pyrochlore lattice can be decomposed into nonoverlapping hexagonal loops -- the hard hexagons -- which exhibit a robust finite-size gap of $\sim 0.69J$.

Second, the hexagons are connected  by a quartet of bonds linking two pairs of antiferromagnetically correlated spins \emph{symmetrically}, making the coupling \emph{doubly frustrated} [\autoref{fig:hard_hexaggon_cover}(d)] \cite{nersesyan_spinons_2003}.
This latter point is perhaps conceptually the most interesting one, as it shifts the perspective from the frustration-induced degeneracy arising on a single tetrahedron to the isolation of \emph{unfrustrated} (nondegenerate) geometrical motifs. 
Third, the \emph{kinetic} energy of local defects in this background pattern is similarly suppressed.
Fourth, there are no matrix elements between different hard-hexagon coverings to any finite order in perturbation theory.

From this point forward, we discuss in detail the $S=1/2$ hard-hexagon state on the pyrochlore lattice. Having understood its essential ingredients, we are able to apply similar constructions to the $S=1$ pyrochlore Heisenberg model as well as Heisenberg models
on the ruby [\autoref{fig:hard_hexaggon_cover}(b)] and checkerboard [\autoref{fig:hard_hexaggon_cover}(c)] lattices. In these cases, we verify that the variational energy of these states is competitive with the ground-state energy obtained by density matrix renormalization group (DMRG) methods ~\cite{Verstraete_matrix_2004,schollwock_the_2005,Feiguin_finite_2005,schollwock_the_2011,stoudenmire_studying_2012} [\autoref{tab:gs_energies}].

We start by covering the pyrochlore lattice with hexagons such that each site participates in exactly one hexagon.
The number of ways to do this is exponentially large in
the linear system size, as we discuss in more detail later.
For a given covering, we then
decompose the nearest-neighbor Heisenberg model into links within the hard hexagons, $H_0$, and the connecting terms, $V$:
\begin{align}\label{eq:hamiltonian}
    H =& H_0 + V \nonumber\\
    =& J\sum_{\langle i,j\rangle\in \hexagon} \vec{S}_i\cdot \vec{S}_j + J\sum_{\langle i,j\rangle\notin\hexagon} \vec{S}_i\cdot \vec{S}_j
\end{align}

\paragraph{Ground-state energy.}
\begin{figure}
	\includegraphics{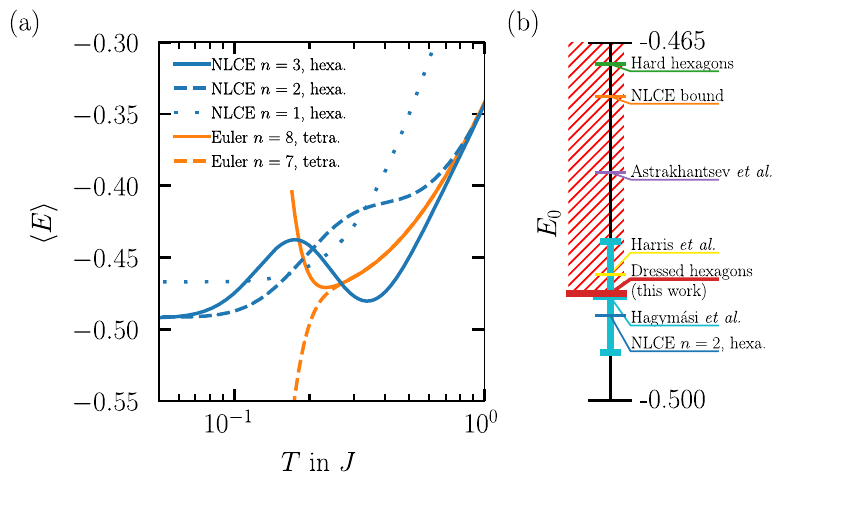}
	\caption{(a) Energy per site at finite temperature for the pyrochlore $S=1/2$ and $J=1$. The orange curves are showing a tetrahedron-based NLCE expansion up to eighth order \cite{schafer_pyrochlore_2020}, which defines an upper limit for the ground-state energy. The blue curves show the NLCE expansion based on the hexagons. {{Euler refers to a resummation algorithm extending the convergence down to lower temperatures \cite{schafer_pyrochlore_2020}}}. (b) Different estimates for the ground-state energy per site: bare hard hexagons (ground state of $H_0$), upper bound obtained by the converged tetrahedron expansion \cite{schafer_pyrochlore_2020}, Astrakhantsev \emph{et al.}  \cite{astrakhantsev_broken-symmety_2021}, Harris \emph{et al.} \cite{harris_ordering_1991}, dressed hexagon  state $E_\alpha$ (this work) [\autoref{eq:psi_alpha}], Hagym\'asi \emph{et al.} \cite{hagymasi_possible_2021} and the NLCE hexagon expansion at second order for $T=0$ (this work). Estimates of the ground-state energy higher than $E_{\alpha}$ the variational energy can be ruled out.}
	\label{fig:pyrochlore_s12_finite_temp}
\end{figure}

Crucially, the decoupled hexagon Hamiltonian $H_0$ exhibits a large gap
$\sim0.69 J$, while the coupling $V$ is effectively suppressed
since it {\it symmetrically} couples  pairs of neighboring--and hence {\it anti}ferromagnetically correlated--spins in two hexagons  via the four bonds [{Fig.~\ref{fig:hard_hexaggon_cover}(d)}] of their shared tetrahedron.
Moreover, the ground-state energy of $H_0$, $E_0 \sim -0.47J$, is already not
far from the pre-existing estimates of the ground-state energy
of the full pyrochlore lattice.
This motivates a variational approach by which we establish
a strict upper bound on the ground-state energy of \autoref{eq:hamiltonian} in the thermodynamic limit which, within error bars, competes with previous extrapolations based on estimates for small clusters \cite{harris_ordering_1991,hagymasi_possible_2021,astrakhantsev_broken-symmety_2021,hagymasi_enhanced_2022}.

The trial wave function is constructed by dressing the ground state of $H_0$, $\ket{\Psi_0}$ -- a simple product state.
To introduce additional correlations between the hexagons and minimize the energy further, we perform imaginary-time evolution using the Hamiltonian of the tetrahedral links, $V$, connecting the hexagons:
\begin{align}
    &\ket{\Psi_\alpha}=e^{-\alpha V}\ket{\Psi_0}\label{eq:psi_alpha}\\
    \Rightarrow\quad & E_\alpha = \frac{1}{N}\frac{\braOket{\Psi_\alpha}{H}{\Psi_\alpha}}{\braket{\Psi_\alpha}{\Psi_\alpha}}
\end{align}
The variational energy per site $E_\alpha$, which we evaluate using an expansion in powers of $\alpha$ \cite{ursell_evaluation_1927,brueckner_two-body_1955,goldstone_derivation_1957,percus_correlation_1975,shlosman_signs_1986,mattuck_guide_1992,supp}, 
exhibits a well-defined minimum at $\alpha=\alpha_0$. 
{The expansion is fully converged around the minimum and the truncation error of the variational energy is much smaller than the symbol size in \autoref{fig:pyrochlore_s12_finite_temp} (b).}
Importantly, the resulting minimal energy $E_{\alpha_0}$ (see \autoref{tab:gs_energies}), due to its variational nature, rules out all larger estimates from previous studies \cite{harris_ordering_1991, astrakhantsev_broken-symmety_2021}, as indicated in \autoref{fig:pyrochlore_s12_finite_temp} (b).

% E0 = -0.48983821830302543 +/- 0.005737698765539114  spin 1/2 from Hagymasi et al.
\setlength{\tabcolsep}{3pt}
\begin{table}[t]
\centering
\begin{tabular}{l | c | c | c }\hline\hline
   model & NLCE$_2$ & DMRG & $E_{\alpha_0}$  \\ \hline   
Pyrochlore, $S=\frac{1}{2}$   &-0.4917(5) & -0.490(6) &   -0.489472(8)  \\
Ruby, $S=\frac{1}{2}$   & -0.492(1) &  -0.4865(7) &  -0.48946(5) \\
Checkerboard, $S=\frac{1}{2}$   & -0.5138(2) &  -0.5132(2) &  -0.513442(1)  \\
Pyrochlore, $S=1$   & -1.489(5) & -1.520(6){$^*$} & -1.490(1) \\
Ruby, $S=1$   &   -1.489(5) &    -1.4764(5) &  -1.490(1) \\
Checkerboard, $S=1$   & -1.533(2) &  -1.532(1) & -1.5339(5)\\\hline\hline 
\end{tabular}
\caption{Ground-state energies in units of $J$ for different models and spin lengths. 
The ground-state energy is calculated using second-order NLCE, DMRG, and the variational wave function \autoref{eq:psi_alpha} with optimization of the parameter $\alpha$. The optimal values of $\alpha$ are given in Tab. S1 the Supplementary Material \cite{supp}.
%In all cases, the optimized variational energy agrees closely with NLCE and is competitive with the energy obtained in DMRG. 
The DMRG energies for the pyrochlore lattice are from Hagym\'asi \emph{et al.} \cite{hagymasi_possible_2021,hagymasi_enhanced_2022}, while the DMRG results were obtained using ITensor\cite{fishman_itensor_2022}. {$^*$Note that the $S=1$ pyrochlore case was obtained for $48$ sites, and finite-size effects are likely to underestimate the ground-state energy.}}
\label{tab:gs_energies}
\end{table}

We further support the stability of the hard-hexagon state by a numerical linked cluster expansion (NLCE)\cite{tang_nlce_2013,schafer_magnetic_2022}.
This method has proven valuable in determining thermodynamic quantities at \emph{finite} temperature in three-dimensional lattices, including the pyrochlore lattice \cite{applegate_nlce_pyrochlore_exp_comp_2012, singh_nlce_tetra_pyrochlore_2012, schafer_pyrochlore_2020}.
The algorithm is in a spirit similar to a high-temperature expansion in the sense that it systematically includes larger clusters to obtain an estimate at a finite temperature.
While most previous works on the pyrochlore exploit the tetrahedral structure (which is powerful at finite temperature), we generalized the expansion based on nonoverlapping hexagons \cite{supp}.
Typically, the limit of convergence of a finite order expansion is clearly identified by the fact that successive orders rapidly diverge from each other to \emph{infinity} below some temperature.
This is the case for the tetrahedral expansion, as shown in \autoref{fig:pyrochlore_s12_finite_temp} (orange lines).
However, the situation is remarkably different for the hexagon-based expansion (blue lines in the same figure).
It is clearly not converged for intermediate temperatures, $0.06J\leq T \leq 0.8J$, \emph{but} {converges} again for $T\rightarrow 0$ yielding a realistic ground-state energy compatible with previous studies \cite{harris_ordering_1991,hagymasi_possible_2021,astrakhantsev_broken-symmety_2021}.
Strikingly, the low-temperature \emph{convergence} is effectively obtained in the second order.
This lends further credence to a particularly simple low-temperature state of weakly dressed hexagons.
The success of the NLCE at $T=0$ is reminiscent of a study of a distorted kagom\'e
lattice \cite{khatami_nlce_pinwheel_kagome_2011}, where a similar approach was used to
support the conclusion of a dimerized ground state.

\paragraph{Hard-hexagon coverings and symmetry breaking.}

\begin{figure}
	\includegraphics{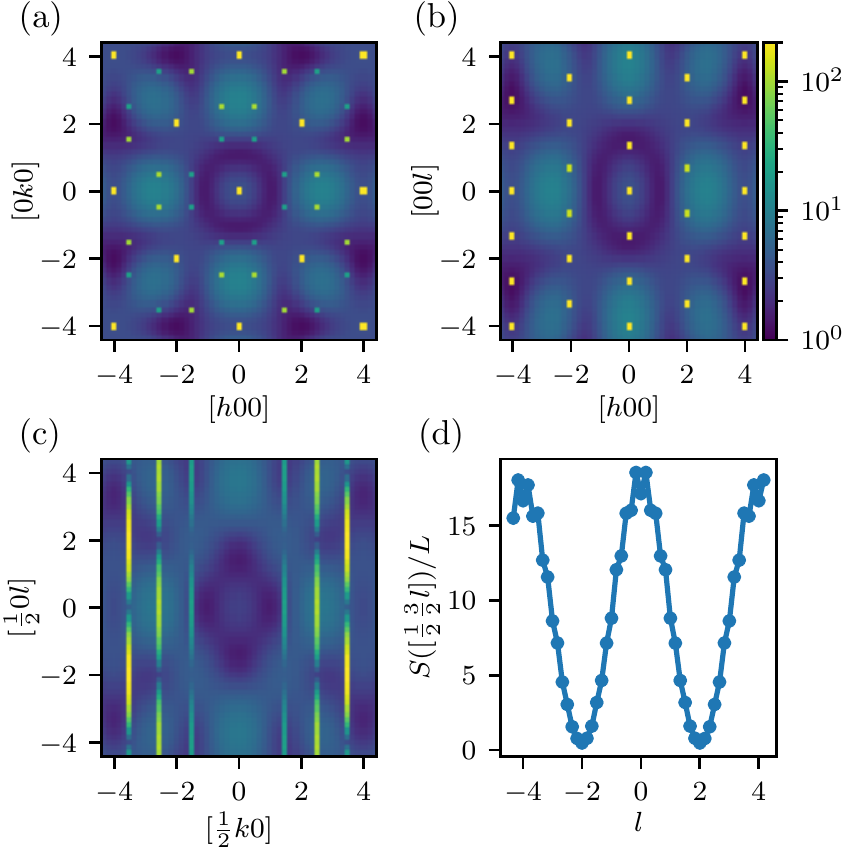}
	\caption{Symmetry breaking of a hard-hexagon state with stacking direction $[001]$. We show the dimer-dimer correlation function [\autoref{eq:dimer_s_q}], which, in contrast to the usual two-point correlation function, shows signatures of both the breaking of translation and rotation symmetry [Bragg peaks in panels (a), (b)] as well as of the disorder in the stacking direction [broad features in panels (c), (d)].}
	\label{fig:dimer-s-q}
\end{figure}

The proposed states are nonmagnetic valence-bond crystals, whose symmetry-breaking properties are determined by the underlying coverings of the lattice by hard hexagons. 
The pyrochlore lattice admits an exponentially large family of hard-hexagon states, all of which take the form of long-ranged ordered planes of hexagons stacked along one of the three equivalent $\langle001\rangle$ directions. Taking for example the covering shown in \autoref{fig:hard_hexaggon_cover} (a)
we can obtain a new valid covering by shifting the second plane from the bottom (in yellow-red) one unit to the right (i.e. along one of the $\langle110\rangle$ crystal directions). 
Assuming that all hard-hexagon coverings of the pyrochlore lattice can be constructed from such shifts, this yields a subextensive, yet exponentially large, number of coverings $N_{\rm cover} = 3 \times 2^{4L/3}$, where $L$ is the linear system size given as the number of cubic, 16-site unit cells. 
We have verified numerically {\cite{supp}}, for finite clusters up to $L=12$, that these are the only coverings by using 
an unbiased numerical optimization algorithm.

The fact that there is an infinite family of states on the pyrochlore lattice, rather than a single hard-hexagon state, is interesting for two main reasons. 
First, for a single such state to be stable, it is important that a finite order of perturbation theory connect no two members of the family. This is clearly the case for the stacked-layer coverings discussed above since going from one to the other requires the translation of an entire plane of hexagons. 
Second, this has implications for the symmetry breaking of a state randomly selected from, or averaged over,  this family. {Since in a valence bond solid, two-point correlation functions decay rapidly,  we instead compute a four-point correlation function which we call the dimer structure factor, or bond correlator,} 
\begin{align}
    S_{\rm dimer}(\vec q) =& \sum_{\expval{ij}, \expval{kl}} \exp(-i \vec q \cdot \left[ \frac{1}{2}(\vec r_i + \vec r_j) - \frac{1}{2}(\vec r_k + \vec r_l) \right]) \nonumber \\ 
    &\times \expval{(\vec S_i \cdot \vec S_j)\, (\vec S_k \cdot \vec S_l)},
    \label{eq:dimer_s_q}
\end{align}
for an ensemble of hard-hexagon states with $[001]$ stacking direction obtained by the numerical optimization algorithm mentioned before. The result is shown in \autoref{fig:dimer-s-q}. In contrast to the ``usual'' spin structure factor based on a two-point correlation function, the dimer structure factor is based on a four-point correlation function and carries signatures of the translational and rotational symmetry breaking [Bragg peaks in panels (a) and (b)] as well as the disorder in the stacking direction [broad features in panels (c) and (d)] and also the short-range correlations within the hexagons (low-intensity broad features rendered visible by  the log scale). 

{We note that, given the non-magnetic nature of the symmetry-breaking, probes such as Raman scattering or ultrasound measurements may be useful experimentally besides neutron scattering, which can best establish the absence of long-range magnetic order.}

\paragraph{Robustness of the gap.}

\begin{figure}
	\includegraphics{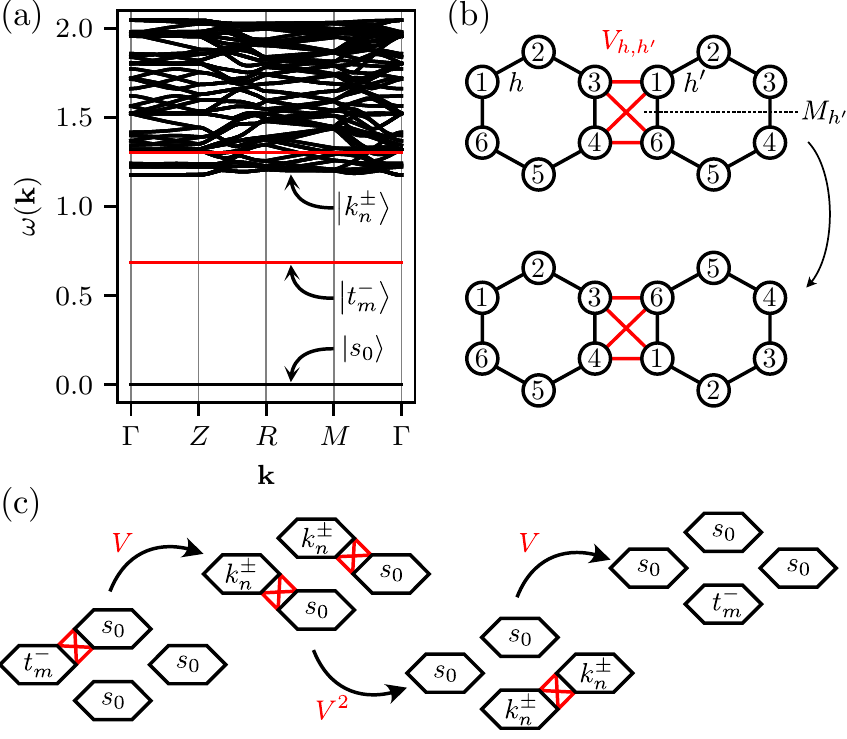}
	\caption{(a) Dispersion of excitations above the hard-hexagon state in \autoref{fig:hard_hexaggon_cover} (a) up to linear order in $V$. (b) Coupling of hexagons and resulting local symmetry of the hopping matrix element $V$. {Flat bands are colored in red for visibility}
 (c) Schematic of the fourth-order hopping process by which the lowest-lying triplet $\ket{t^{-}_m}$ would gain kinetic energy}
	\label{fig:kinetic-energy}
\end{figure}

Since the hard-hexagon states break only discrete symmetries, they are -- if stable -- generically gapped. We analyze the excitations above our family of hexagon states to asses their (local) stability.
As the candidate ground states are weakly dressed product states of single-hexagon ground states, the simplest ansatz for the excited states is given analogously by weakly dressed local triplet excitations on a single hexagon. The lowest-lying excited state on a single hexagon is given by a triplet, with a gap of $\sim 0.69J$.
A possibility to reduce and finally destabilize the gap would be for the localized excitations to gain \emph{kinetic} energy. 
{Numerically evaluating the excitation dispersion in three dimensions is difficult, and presently out of reach. Instead, we turn to an analytic approximation: namely the multiboson expansion, which is able to describe the hopping of local excitations above a given ground state, at first order in perturbation theory \cite{sachdev1990, romhanyi2012, supp}.} The results are shown in \autoref{fig:kinetic-energy}: remarkably, the lowest-lying triplet bands, $\ket{t_m^-}$ ($m=-1, 0, 1$) at $E\sim 0.69J$, remain completely flat, while the higher energy triplets disperse (bands labeled by $\ket{k_n^\pm}$ in \autoref{fig:kinetic-energy}) and are actually pushed below the excited singlet located at $E\sim 1.3J$, which also remains flat. 
The flatness of the bands is due to a symmetry of the hopping matrix elements $V$, \cf see \autoref{eq:hamiltonian}, connecting two hexagons $A$ and $B$. This is illustrated in \autoref{fig:kinetic-energy} (b): mirroring each hexagon \emph{individually} across a plane running through the tetrahedron leaves the operator matrix element $V(A, B)$ invariant.
This implies a selection rule for the hopping matrix element $\bra{s_0, t} V \ket{t, s_0}$.
Classifying all eigenstates of $H_0$ by the three mirror symmetries of a hexagon, it turns out that the ground state singlet $\ket{s_0}$ as well as the lowest-lying triplet $\ket{t_m^-}$ are simultaneous eigenstates of all three, but with eigenvalue $+1$ and $-1$ respectively. 
This then guarantees $\bra{s_0, t_m^-} V \ket{t_m^-, s_0}=0$ by symmetry, hence the flatness of the lowest band.
By contrast, the two triplet pairs at $E\sim 1.5J$ and $E\sim 1.8J$  are \emph{not} simultaneous eigenstates of all three mirror operations and hence are not prevented from  mixing into dispersive modes.
In \autoref{fig:kinetic-energy} (c), we show schematically a process with several high-energy virtual intermediate states, which would give the lowest-lying triplet kinetic energy. However, since this only appears at the fourth order in $V$,  we do not expect this to lower the gap of the hard-hexagon state significantly. 
Finally, we note that the symmetry argument above does not depend on the choice of a particular tiling, but the exact shape of the dispersion of high-energy bands in \autoref{fig:kinetic-energy} (a) will. 

To summarize our results for the pyrochlore $S=1/2$ model:
we propose an exponentially large family of valence-bond crystals based on hard-hexagon coverings as candidate ground states. A variational calculation of their energy in the thermodynamic limit is within the error bars of the best numerical estimates of the ground-state energy. The stability of these states is further supported by NLCE and by a multiboson expansion which shows that the kinetic energy of excitations is suppressed.
The hard-hexagon coverings break rotation, translation, and inversion symmetries of the lattice.
It is important to note that the recent numerical studies that found lattice symmetry breaking \cite{hagymasi_possible_2021, astrakhantsev_broken-symmety_2021} used clusters that are incompatible with the hard-hexagon states, rendering them energetically disfavorable there.

{Note that the same ingredients rendering the stability of the valance bond state are found in other Heisenberg models, such as the two-dimensional ruby and checkerboard lattice for $S=1/2$ and $S=1$, \cf \autoref{fig:hard_hexaggon_cover}. We obtain analogous results, such as the flatness of the triplet band and a finite size gap, in these cases. 
A short discussion of spin $S=1$~\cite{haldane_nonlinear_1983} and broken symmetries~\cite{shastry1981,sindzingre02, fouet_planar_2003,brenig_planar_2002, chan2011, bishop2012} are given in the supplement.}

\paragraph{Discussion.} Having identified a new family of energetically competitive states based on effectively decoupled close-packed motifs, we still cannot settle the question of what is the actual ground state in the pyrochlore case. On a technical level, our study underlines the need to consider possibly very large unit cells in finite-size studies.  
{We note that such a change in perspective was already proposed in the context of the  $S=3/2$ spinel compound ZnCr$_2$O$_4$ \cite{lee_emergent_2002}, where hexagonal motifs had been identified in a study of its magnetoelastic properties \cite{tchernyshyov_order_2002}. 
In this case, theoretical modeling ultimately suggested that the system is not described by independent hexagonal clusters but rather that weak further neighbor exchange can account for the salient observations \cite{conlon_absent_2010,Fennell_hex_2008}.
In this work, we have demonstrated that the formation of such clusters can indeed occur, but via a different, quantum, mechanism.}

{The existence of potential material realizations of the pyrochlore Heisenberg antiferromagnet \cite{clark_pyrochlore_s1_exp_2014, ShuZhang_NCNF_dyn}, and the growing capabilities of cold atom emulations 
\cite{zhang18, browaeys20, semeghini21}, gives hope that some
of these questions may eventually be settled by experiment.
The presence of a significant gap to spin excitations above the 
hard hexagon states would be an important signature for both thermodynamic and spectroscopy measurements if indeed the system
can equilibrate into such a state at low temperatures.
The presence of sharp, gapped, $S=1$ excitations, along with the presence of lattice symmetry breaking would serve to distinguish the hard hexagon states from competing quantum spin liquids.
}

{
However, due to the presence of such a large family of 
low energy states, as well as potential additional states
beyond the ansatz considered here, it may be that
the physics at the temperatures reachable in  the experiment is controlled not by a single ground state but by many competing ones}. Further, the eventual ground-state selection in an actual material in the presence of any near degeneracies will take place via any residual deviations from an ideal Heisenberg Hamiltonian, and may require exquisitely low temperatures. Finally, hexagonal motifs have appeared in various pyrochlore settings \cite{tchernyshyov_order_2002,conlon_absent_2010,Fennell_hex_2008}, and the mechanisms leading to their stabilization may reinforce each other in a given material.

\begin{acknowledgments}
    We are very grateful to Imre Hagym\'asi, Christopher Laumann, David J. Luitz, Frank Pollmann, Götz S. Uhrig, and Alexander Wietek for many helpful discussions on this topic.
    DMRG and TDVP calculations were performed using the ITensor\cite{fishman_itensor_2022} package with the global subspace expansion\cite{mingru_time-dependent_2020}.
	This work was in part supported by the Deutsche Forschungsgemeinschaft under Grant SFB 1143 (Project-ID No. 247310070) and the cluster of excellence 
ct.qmat (EXC 2147, Project-ID No. 390858490).
    RS was further supported by AFOSR Grant No. FA9550-20-1-0235.
\end{acknowledgments}

R.S. and B.P. contributed equally to this work.

\bibliography{pyrochlore}

\end{document}

\title{Supplementary Material\\Abundance of hard-hexagon crystals in the quantum pyrochlore %Heisenberg
antiferromagnet}
\author{Robin Sch\"afer}\email{schaefer@pks.mpg.de}
\author{Benedikt Placke}\email{placke@pks.mpg.de}
\author{Owen Benton}\email{benton@pks.mpg.de}
\author{Roderich Moessner}\email{moessner@pks.mpg.de}
\affiliation{Max Planck Institute for the Physics of Complex Systems, Noethnitzer Str. 38, 01187 Dresden, Germany}

\date{\today}

\begin{abstract}
In this supplementary material, we provide additional details of the numerical calculations and analytical arguments given in our letter.
\end{abstract} 

\maketitle
\tableofcontents

\section{Variational state\label{app:time_evol}}

\begin{figure}[t]
    \centering
    \includegraphics[width=0.66\columnwidth]{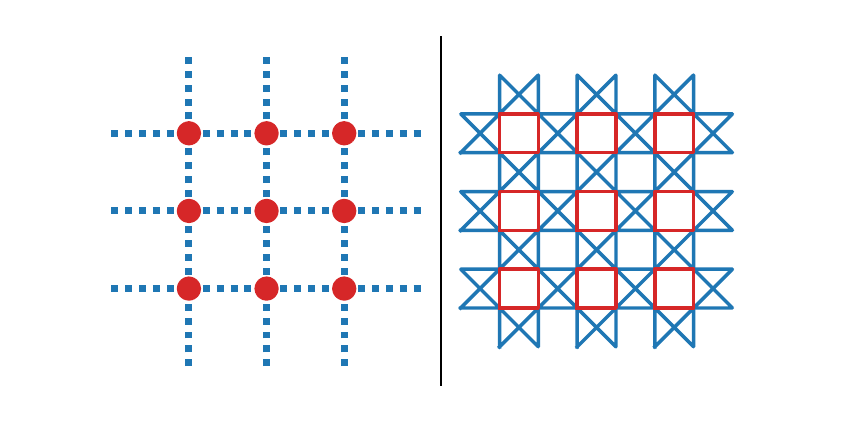}
    \caption{Reduced lattice (left) for the checkerboard lattice (right). We identify the plaquettes with the nodes and the links of the tetrahedron connecting two plaquettes with the links of the reduced lattice. The same decomposition was exploited for the NLCE algorithm.}
    \label{fig:lattice_checkerboard}
\end{figure}

The family of valence-bond crystals that we propose in the main text is applicable to Hamiltonians which can be decomposed into \emph{plaquette} terms and \emph{edge} terms connecting two plaquettes:
\begin{align}
    H_0 = \sum_p H_p \quad\text{and}\quad V = \sum_e V_e.
\end{align}
In the case of the pyrochlore lattice, the ``plaquettes'' correspond to the hexagons and $V_e$ refer to the quartet of bonds connecting two hexagons. Another example -- the checkerboard lattice -- is shown on the right of \autoref{fig:lattice_checkerboard}.

An obvious starting point for building a trial wave function for our family of states is the product state of single-hexagon ground states $\ket{s_0}$
\begin{equation}
    \ket{\Psi_0} = \bigotimes_p \ket{s_0}_p.
\end{equation}
Which is an eigenstate of $H_0$ and the zero expectation value vanishes for the perturbation $V$ in our case. Hence, it has an energy expectation value $E_0 = \expval{H}{\Psi_0} = \expval{H_0}{\Psi_0}$ ($\sim -0.47J$ in the pyrochlore for example).
To make the energy of $\ket{\Psi_0}$ competitive with values in the literature\cite{fouet_planar_2003,brenig_planar_2002,hagymasi_possible_2021,astrakhantsev_broken-symmety_2021}, we have to weakly \emph{dress} it. 
This could in principle be done perturbatively. However, because of the simplicity of our proposed family of states, we are able to construct a dressed variational \emph{wave function}, which has the great advantage of yielding a strict upper bound on the ground-state energy:
\begin{equation}
    \ket{\Psi_{\vec v}} = e^{-S(\vec v)} \ket{\Psi_0}
\end{equation} 
where $\vec v$ is some set of variational parameters and $S(\vec v)$ some local operator. 
In this paper, we consider a very simple form $S=\alpha V$ with only a single variational parameter $\alpha$. It can be interpreted as an imaginary-time evolution of $\ket{\Psi_0}$ under $V$ by a time $\alpha \in\mathbb{R}$
\begin{equation}\label{eq:psi_alpha}
    \ket{\Psi_{\alpha}} = e^{-\alpha V} \ket{\Psi_0}.
\end{equation} 
While $\alpha=0$ corresponds to just the single-hexagon product state $\ket{\Psi_0}$, increasing $\alpha$ constitutes a trade-off between paying energy on the hexagons to gain some energy on the connecting bonds.

The variational energy per site is then given by
\begin{align}
    E_\alpha &= \frac{1}{L_p}\frac{1}{N_p}\frac{\braOket{\Psi_\alpha}{H_0+V}{\Psi_\alpha}}{\braket{\Psi_\alpha}{\Psi_\alpha}}=\frac{1}{L_p}\frac{1}{N_p}\frac{\braOket{\Psi_0}{e^{-\alpha V}(H_0+V)e^{-\alpha V}}{\Psi_0}}{\braOket{\Psi_0}{e^{-2\alpha V}}{\Psi_0}}\label{eq:E_a}
\end{align}
where $N_p$ is the number of plaquettes and $L_p$ is the length of the single plaquette.

\subsection{Linked Cluster Theorem}
The exponential form of the variational state \autoref{eq:psi_alpha} makes it possible to develop a systematic linked cluster expansion\cite{brueckner_two-body_1955,goldstone_derivation_1957,mattuck_guide_1992} to evaluate the trial \emph{energy} $E_\alpha$ in \autoref{eq:E_a} in the thermodynamic limit. In the remainder of this section, we derive this expansion, explain how it is implemented and show that it reveals a pronounced minimum in the variational energy. 
We emphasize that the linked cluster
expansion used here to calculate the variational energy is different to the numerical linked cluster expansion described in the main text and in \autoref{app:NLCE}, which is based on exact diagonalization and does not require a variational ansatz.

To start, note that $\ket{\Psi_\alpha}$ is a spin-singlet (as it should be for a $SU(2)$ symmetric Hamiltonian), and all plaquettes, as well as the edge terms connecting two plaquettes, are equivalent, respectively. Hence, to compute the variational energy $E_\alpha$, it suffices to compute the energy of a single plaquette term and a single connecting term:
\begin{align}
    E_\alpha = \frac{1}{L_p}\left(\langle H_p \rangle_\alpha + \frac{c}{2}\langle V_e\rangle_\alpha\right)
\end{align}
where $\langle O \rangle_\alpha = \expval{O}{\Psi_\alpha}$ for some operator $O$ and $c$ is the number of edge terms $V_e$ attached to each plaquette. In the cases considered throughout this letter, it is given by the length of the plaquette.

Following the linked cluster theorem, the expectation value of any local operator can be written in terms of connected correlation functions, also known as \emph{Ursell} functions \cite{ursell_evaluation_1927,percus_correlation_1975,shlosman_signs_1986}. 
To obtain the expectation value of some local operator $O_{\text{loc}}$ we expand the exponentials in $e^{-\alpha V}O_{\text{loc}}e^{-\alpha V}$ and only consider terms that are connected to the local operator.
Each connected term with $n$ operators is weighted according to the $n$th Ursell function; the first two Ursell functions are given by 
\begin{align}
     u_{\ket{\Psi}}\left[A\right]=& \braOket{\Psi}{A}{\Psi}\label{eq:ursell1}\\
     u_{\ket{\Psi}}\left[A,B\right] =&\braOket{\Psi}{AB}{\Psi} -\braOket{\Psi}{A}{\Psi}\braOket{\Psi}{B}{\Psi}. \label{eq:ursell2}
\end{align}

{{
The linked cluster theorem~\cite{brueckner_two-body_1955,goldstone_derivation_1957,mattuck_guide_1992} states that non-connected terms that factorize do not contribute to the local observable.
In the technical sense, connected terms are those with connected support given by the operators stemming from the exponentials.
}}
This yields 
\begin{align}\label{eq:linked_cluster}
    \langle O_{\text{loc}}\rangle_\alpha=&\frac{\braOket{\Psi_\alpha}{O_{\text{loc}}}{\Psi_\alpha}}{\braket{\Psi_\alpha}{\Psi_\alpha}}
    &= \sum_{a,b=0}^\infty \sum_{\substack{{e_1,\dots,e_{a+b}}\\\text{ connected to }O_{\text{loc}}}} \frac{(-\alpha)^{a+b}}{a!b!}u_{\ket{\Psi}}\left[V_{e_1},\dots, V_{e_a},O_{\text{loc}},V_{e_{a+1}},\dots,V_{e_{a+b}}\right]
\end{align}
In order to identify a set of edges $\{e_1,\dots,e_n\}$ ``connected'' to $O_{\rm loc}$, it is useful to introduce the notion of the \emph{reduced lattice}. The reduced lattice for the partition of plaquettes in the checkerboard lattice is shown in \autoref{fig:lattice_checkerboard}. Each vertex $v$ in the reduced lattice corresponds to a plaquette, and two vertices are connected by an edge $e=(v, v^\prime)$ if the two corresponding plaquettes, $v$ and $v^\prime$, are connected. 
A set of edges $\{e_i\}$ in this language is then ``connected'' to $O_{\rm loc}$ if the subgraph induced by the $\{e_i\}$, joint with the support of $O_{\rm loc}$, is connected. 

We illustrate \autoref{eq:linked_cluster} by evaluating $\langle O_{\text{loc}} \rangle_\alpha$ to first order in $\alpha$.
The nominator and denominator in \autoref{eq:linked_cluster} have the following forms:

\begin{align}
    \braOket{\Psi_\alpha}{O_{\text{loc}}}{\Psi_\alpha} =& \braOket{\Psi_0}{\left(1-\alpha V +\mathcal{O}(\alpha^2)\right)O_{\text{loc}}\left(1-\alpha V  +\mathcal{O}(\alpha^2)\right) }{\Psi_0}\label{eq:order1_A} \\
    =& \braOket{\Psi_0}{O_{\text{loc}}}{\Psi_0}-2\alpha \sum_e\braOket{\Psi_0}{V_eO_\text{loc}}{\Psi_0} +\mathcal{O}(\alpha^2)\nonumber\\    =&\braOket{\Psi_0}{O_{\text{loc}}}{\Psi_0}-2\alpha \sum_{e\in\mathcal{N}\left(O_{\text{loc}}\right)}\braOket{\Psi_0}{V_eO_\text{loc}}{\Psi_0}-2\alpha \sum_{e\notin\mathcal{N}\left(O_{\text{loc}}\right)}\braOket{\Psi_0}{V_e}{\Psi_0}\braOket{\Psi_0}{O_\text{loc}}{\Psi_0}+\mathcal{O}(\alpha^2)\nonumber\\
    \braket{\Psi_\alpha}{\Psi_\alpha} =& \braOket{\Psi_0}{1-2\alpha V +\mathcal{O}\left(\alpha^2\right)}{\Psi_0} 
    = 1 -2\alpha \sum_e\braOket{\Psi_0}{V_e}{\Psi_0}+\mathcal{O}(\alpha^2)\label{eq:order1_B}
\end{align}
Here, $\mathcal{N}(O_{\text{loc}})$ contains ``edges'' $e$ such that $V_e$ and $O_{\text{loc}}$ have overlapping support. For example, a plaquette term $H_p$ in the checkerboard lattice exhibits four neighboring edges $V_e$ each containing a quartet of doubly frustrated bonds.
We have used that all operators are Hermitian and the expectation value of two operators with no common support factorizes
\begin{align}
       \braOket{\Psi}{AB}{\Psi}=\braOket{\Psi}{A}{\Psi}\braOket{\Psi}{B}{\Psi}\quad \text{for}\quad B \notin \mathcal{N}(A).
\end{align}
Evaluating \autoref{eq:linked_cluster} according to the linked cluster theorem is restricted to neighboring terms of $O_{\text{loc}}$ weighed with the Ursell functions which yields
 \begin{align}
     \langle O_{\text{loc}}\rangle_\alpha  =&u_{\ket{\Psi_0}}\left[{O_{\text{loc}}}\right] -\alpha \sum_{e\in\mathcal{N}\left(O_{\text{loc}}\right)}u_{\ket{\Psi_0}}\left[{V_e,O_{\text{loc}}}\right]-\alpha \sum_{e\in\mathcal{N}\left(O_{\text{loc}}\right)}u_{\ket{\Psi_0}}\left[{O_{\text{loc}}},V_e\right]+\mathcal{O}\left(\alpha^2\right)\label{eq:order1_C}\\
     =&\braOket{\Psi_0}{O_{\text{loc}}}{\Psi_0} -2\alpha \sum_{e\in\mathcal{N}\left(O_{\text{loc}}\right)}\braOket{\Psi_0}{V_eO_{\text{loc}}}{\Psi_0}+2\alpha\sum_{e\in\mathcal{N}\left(O_{\text{loc}}\right)}\braOket{\Psi_0}{V_e}{\Psi_0}\braOket{\Psi_0}{O_{\text{loc}}}{\Psi_0}+\mathcal{O}\left(\alpha^2\right).\nonumber%\\
 \end{align}

Collecting the terms in from \autoref{eq:order1_A}, \autoref{eq:order1_B} and \autoref{eq:order1_C} confirms the linked clusters theorem to first order in $\alpha$ as stated in \autoref{eq:linked_cluster}:
\begin{align}
    \langle O_{\text{loc}}\rangle_\alpha \cdot \braket{\Psi_\alpha}{\Psi_\alpha}  =  \braOket{\Psi_\alpha}{O_{\text{loc}}}{\Psi_\alpha}+\mathcal{O}\left(\alpha^2\right)
\end{align}

Analogously, it can be shown that this procedure holds for all powers of $\alpha$ \cite{goldstone_derivation_1957}.
However, the possibility to carry out the expansion straightforwardly is quickly exhausted as the complexity scales super exponentially. 
Therefore, in the next section we present a systematic way to determine the variational energy.

\begin{figure}[t]
     \centering
     \begin{subfigure}[b]{0.49\textwidth}
         \centering
         \includegraphics[width=\textwidth]{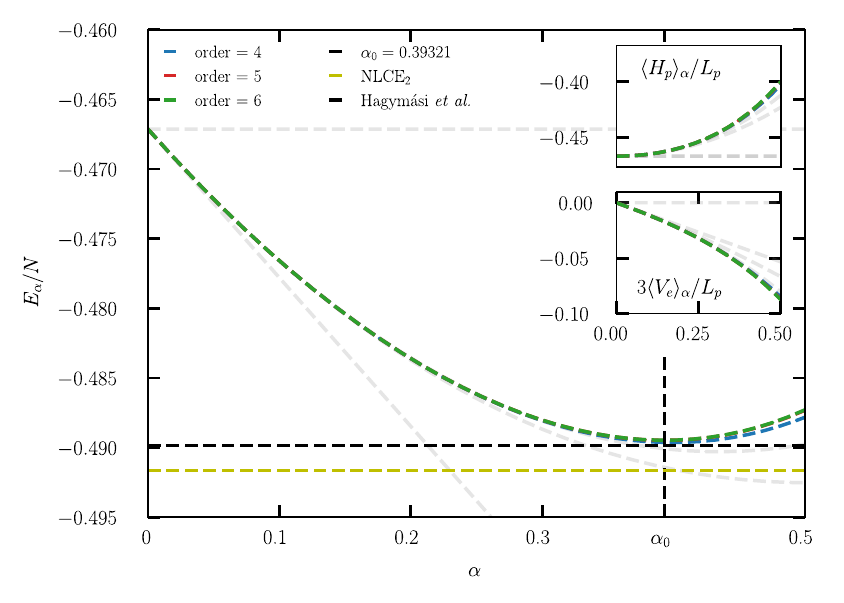}
         \centering
         \caption{Pyrochlore, $S=1/2$.}
         \label{fig:pyrochlore_M=2_E_alpha_a}
     \end{subfigure}
     \hfill
     \begin{subfigure}[b]{0.49\textwidth}
         \centering
         \includegraphics[width=\textwidth]{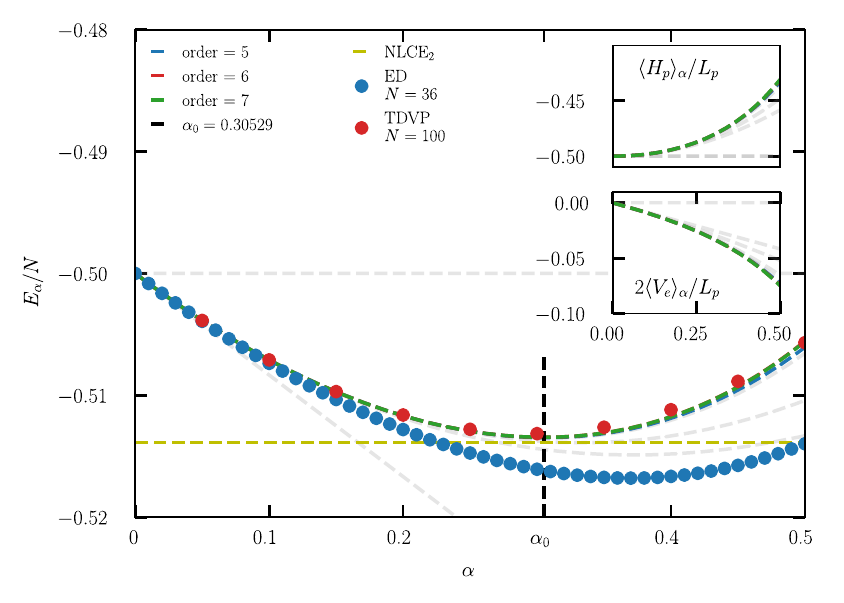}
         \caption{Checkerboard, $S=1/2$.}
         \label{fig:tdvp_checkerboard}
     \end{subfigure}
        \caption{(a) Evaluation of the variational energy $E_\alpha$ in the pyrochlore $S=1/2$. 
        Above 4th order of the linked cluster expansion
        the series is converged over the region containing the minimum at $\alpha=\alpha_0$. The different lines show the result at different orders of the linked cluster expansion are indistinguishable over much of the plot, as a result of the high degree of convergence.
        The inset shows the respective contribution of $\langle H_p\rangle_\alpha$ and $\langle V_e\rangle_\alpha$. Gray lines indicate lower orders of expansion. The sixth-order calculation of $\langle V_e\rangle_\alpha$ omitted the largest clusters containing six and seven hexagons for computational reasons. The contribution of the diagrams in this order is extremely small, and we do not expect this to affect the result. We compare the results to second-order NLCE and the DMRG results of Hagym\'asi \emph{et al.} \cite{hagymasi_possible_2021}. (b) The same calculation was performed for the checkerboard lattice and its comparison to ED and TDVP.
        ED calculations are conducted using the finite-temperature Lanczos method \cite{schnack_magnetism_2018}, 
        with $2 \alpha$ playing the role of inverse temperature.
        While the ED results ($N=36$) have significant finite-size effects the TDVP calculation ($N=100$) with bond dimension $\chi=2000$ agrees nicely with the  expansion of the variational energy, \cf \autoref{app:mps}.}
        \label{fig:pyrochlore_M=2_E_alpha}
\end{figure}

\subsection{Computing the Expansion}
Evaluating \autoref{eq:linked_cluster} requires generating all possible configurations of connected edge terms contained in the sum for a given $a$ and $b$ with  $a+b=n\geq 1$.
{{The basic recipe to evaluate a single term in the sum over $a$ and $b$ can be divided into three steps: (a) determining all connected clusters (spanned by the operators $V_e$ and $O_{\text{loc}}$), (b) identifying all contributing diagrams obtained from these connected clusters, and (c) evaluating all diagrams:}}
\begin{itemize}
    \item[(a)] For $k=1,\dots,n$ we generate the sets $\mathcal{C}_k$ containing $k$ different edges that are, together with $O_{\text{loc}}$, connected. The expansions for $H_p$ and $V_e$ are different.
    \item[(b)] Given a specific set with $k$ edges $\{q_1,\dots,q_k\}\in\mathcal{C}_k$, we have to generate all possible (ordered) sequences of $n$ edges $\left({e_1},\dots,{e_n}\right)$ such that each $q_i$ occurs at least once.
    The restriction that each $q_i$ has to occur at least once ensures the connectivity together with the local operator.
    We denote the set of all generated edge sequences by $\mathcal{S}$.
    \item[(c)] Each edge sequence $\left({e_1},\dots,{e_n}\right)\in\mathcal{S}$ generates an operator string composed of $n+1$ components where the local operator is inserted at position $a+1$. 
    This corresponds to an individual term in \autoref{eq:linked_cluster} which is given by the $(n+1)th$ Ursell function and evaluates to $u_{\ket{\Psi_0}}\left[V_{e_1}\dots V_{e_a} O_{\text{loc}}V_{e_{a+1}}\dots V_{e_{a+b}}\right]$. %, and is weighted according to the $(n+1)$th 
\end{itemize}
Pushing the expansion to high order is challenging as each individual step described above scales \emph{exponentially} or even \emph{factorially} with $n$. 
Luckily, in our case, the above procedure simplifies significantly since the expectation value of any single $V_e$ with respect to a product state on the two plaquettes vanishes if one of the two states is the ground state singlet
\begin{equation}
    \left(\bra{s_0}\otimes\bra{\psi}\right) \,V_e \,\left(\ket{s_0}\otimes\ket{\psi}\right) = 0,~\forall \ket{\psi}.
\end{equation}
This implies that the expectation value of a product of $V_e$ with respect to $\ket{\Psi_0}$ vanishes unless each plaquette is touched by at least \emph{two} $V_e$. Referring to the recipe above, this means that in step (b), we only need to generate sequences such that each vertex is either part of at least two edges or lies in support of the local operator $O_{\rm loc}$.

\setlength{\tabcolsep}{8pt}
\begin{table}[t]
\centering
\begin{tabular}{ c c c c c c}\hline\hline
    \multicolumn{6}{c}{$S=1/2$}\\
    Bethe ($n=8$) & Bethe ($n=6$) &  pyrochlore &  ruby &  Bethe ($n=4$) &  checkerboard  \\   
 0.43(3)  & 0.3961(4) & 0.3932(2) & 0.393(1) & 0.30529(4) & 0.30529(4) \\\hline\hline 
    \multicolumn{6}{c}{$S=1$}\\
    Bethe ($n=8$) & Bethe ($n=6$) &  pyrochlore &  ruby &  Bethe ($n=4$) &  checkerboard  \\   
    0.44(9) &  0.24(7) & 0.24(7) &  0.24(7) & 0.194(4) & 0.197(2) \\\hline\hline 
\end{tabular}
\caption{{Optimal imaginary time step $\alpha_0$ to minimize the energy of the variational wavefunction $\ket{\Psi_{\alpha}}$ for different lattices and spin lengths. The table demonstrates that the underlying lattice topology has negligible influence, as the optimal parameter is mainly determined by the loop length and spin. The error is defined by half the difference between the largest and second largest order. Note that the error is larger than the optimal energy itself (\cf Tab. 1 in the main text) because the minimum is flat and derivations from the optimal $\alpha_0$ have only a marginal influence.}}
\label{tab:alpha0}
\end{table}

\subsection{Results}
The final results for the variational energy $E_\alpha$ [\autoref{eq:E_a}] on the pyrochlore and checkerboard lattice are shown in \autoref{fig:pyrochlore_M=2_E_alpha}. We compute the variational energy using the linked cluster expansion derived above up to order six (seven) for the pyrochlore (checkerboard) lattice. 
{{The order refers to the number of edge terms $V_e$ included in the expansion.}}
As a function of $\alpha$, it shows a well-pronounced local minimum at $\alpha_0$. {The optimal imaginary time steps for the different models considered here are listed in \autoref{tab:alpha0}.} Note that in the relevant range of $\alpha$, orders larger than four (five) are barely distinguishable by the eye in \autoref{fig:pyrochlore_M=2_E_alpha}.
{{At the minimum, the error in the pyrochlore lattice estimated by the difference between order five and six is of order $10^{-4}\,J$ reflecting the convergence of the algorithm. We define the error of the expansion by the difference between the largest and second-largest order at the optimal value $\alpha_0$. This is shown in Tab. 1 in the main text.}}
For the checkerboard lattice, we also perform the time evolution on a large finite cluster of size $N=100$ with periodic boundary conditions using the time-dependent variational principle (TDVP) to check consistency with our linked cluster expansion, finding good agreement (see also \autoref{app:mps}).
Smaller clusters additionally allow the application of an exact finite-temperature Lanczos calculation \cite{schnack_magnetism_2018} starting from an initial plaquette product state where $2\alpha$ serves as the inverse temperature and $V$ as the Hamiltonian.
Implementing this on a finite cluster of the checkerboard lattice with $N=36$, we find the same qualitative dependence of $E_\alpha$ on $\alpha$. However, the result does not agree quantitatively with the other two methods, which we attribute to the presence of strong finite-size effects.

The coefficients of the expansion of the variational energy in powers of $\alpha$ are listed in \autoref{tab:coefficients}. We list not only the pyrochlore and checkerboard but also results for the ruby lattice as well for different \emph{Bethe} lattices. 
By ``Bethe ($n=4, 6, 8$)'' we here denote that the \emph{reduced lattice} is a Bethe lattice. The number $n$ given in parenthesis is the size of the unfrustrated loops constituting $H_0$. These loops are then connected via a quartet of bonds similar to the pyrochlore, checkerboard, and ruby case, but in such a way that the resulting reduced lattice is a regular tree of degree $n$.

We note that for some of the models (marked by an asterix in table \autoref{tab:coefficients}), the coefficients of $\expval{V_e}_\alpha$ at the largest order in the expansion are computed only approximately. Each order in $\alpha$ contains contributions of operator strings with support on clusters of at least one (two) and at most $n+1$ ($n+2$) plaquettes for $O_{\text{loc}}=H_p$ ($O_{\text{loc}}=V_e$).
Due to a large number of arising terms we omitted, for some coefficients of $O_{\text{loc}}=V_e$, the contributions of the largest clusters. We expect this to not make any difference in practice for the following reasons. (i) Most importantly, the expansion is already converged at much lower orders, and (ii) also the contribution $\expval{V_e}_\alpha$ is small compared to that of $\expval{H_p}_\alpha$. (iii) The omitted terms are only a small fraction of those contributing at the given order, and they also typically exhibit small weight.
(iv) The omitted cluster occurs at high order, which is suppressed for $\alpha_0\approx 0.4$. 
All this is also evidenced by the excellent quantitative agreement between the numerical imaginary-time evolution and the linked cluster expansion.

Remarkably, the lattice geometry and dimensionality have little impact on the numerical values of the coefficients in the linked cluster expansion as long as the plaquette length stays fixed. This is true even at orders where the lattices \emph{are} topologically distinct (\eg orders larger than three for the ruby and pyrochlore lattices).
This again validates the (local) stability of the hard-hexagon crystals, as it implies that even closed loops of plaquettes, contributing at higher orders, are negligible.

\setlength{\tabcolsep}{4pt}

\begin{table*}[t]
\centering
\begin{tabular}{ l  | c c c c c c c c  }
\multicolumn{1}{l}{coefficients of $E_{\rm strong} = \expval{H_p}_\alpha/L_p$} \\
\hline\hline 
\diagbox[dir=NW,innerrightsep=0pt,innerleftsep=0pt]{$\qquad\quad$Model$\qquad$}{$\qquad$Order$\qquad\quad$} & $0$ & $1$ & $2$ & $3$ & $4$ & $5$ & $6$ & $7$ \\\hline
Checkerboard, $S=1/2$   &    -0.50000  &    0.00000  &    0.16667  &    -0.08333  &    0.18461  &    -0.11343  &    0.10520  &    -0.05644  \\ 
Bethe ($n=4$), $S=1/2$   &    -0.50000  &    0.00000  &    0.16667  &    -0.08333  &    0.18461  &    -0.11343  &    0.10376  &    -0.05440  \\\hline 
Checkerboard, $S=1$   &    -1.50000  &    0.00000  &    1.00000  &    -0.50000  &    3.02500  &    -2.34290  &    8.45032  &    --  \\ 
Bethe ($n=4$), $S=1$   &    -1.50000  &    0.00000  &    1.00000  &    -0.50000  &    3.02500  &    -2.34290  &    8.41575  &    --  \\\hline 
Pyrochlore, $S=1/2$   &    -0.46713  &    0.00000  &    0.17618  &    -0.08809  &    0.14804  &    -0.08124  &    0.03410  &    --  \\ 
Ruby, $S=1/2$   &    -0.46713  &    0.00000  &    0.17618  &    -0.08809  &    0.14561  &    -0.08292  &    0.01199  &    --  \\
Bethe ($n=6$), $S=1/2$   &    -0.46713  &    0.00000  &    0.17618  &    -0.08809  &    0.14699  &    -0.07885  &    0.02905  &    --  \\\hline 
Pyrochlore, $S=1$   &    -1.43624  &    0.00000  &    1.07960  &    -0.53980  &    2.71700  &    --  &    --  &    --  \\  
Ruby, $S=1$   &    -1.43624  &    0.00000  &    1.07960  &    -0.53980  &    2.69478  &    --  &    --  &    --  \\
Bethe ($n=6$), $S=1$   &    -1.43624  &    0.00000  &    1.07960  &    -0.53980  &    2.69158  &    --  &    --  &    --  \\\hline 
Bethe ($n=8$), $S=1/2$   &    -0.45639  &    0.00000  &    0.17867  &    -0.08933  &    0.13663  &    --  &    --  &    --  \\\hline  
Bethe ($n=8$), $S=1$   &    -1.41712  &    0.00000  &    1.10135  &    -0.55067  &    --  &    --  &    --  &    --  \\\hline\hline  
\multicolumn{8}{c}{} \\
\multicolumn{1}{l}{coefficients of $E_{\rm weak} = \expval{V_e}_\alpha/4$} \\\hline\hline 
\diagbox[dir=NW,innerrightsep=0pt,innerleftsep=0pt]{$\qquad\quad$Model$\qquad$}{$\qquad$Order$\qquad\quad$} & $0$ & $1$ & $2$ & $3$ & $4$ & $5$ & $6$ & $7$ \\\hline
Checkerboard, $S=1/2$   &    0.00000  &    0.04167  &    -0.02083  &    0.05324  &    -0.03819  &    0.04884  &    -0.03658  &    0.03050  \\
Bethe ($n=4$), $S=1/2$   &    0.00000  &    0.04167  &    -0.02083  &    0.05324  &    -0.03819  &    0.04853  &    -0.03609  &    0.02622  \\\hline 
Checkerboard, $S=1$   &    0.00000  &    0.16667  &    -0.08333  &    0.51111  &    -0.44753  &    1.68815  &    -1.87512$^*$  &    --  \\ 
Bethe ($n=4$), $S=1$   &    0.00000  &    0.16667  &    -0.08333  &    0.51111  &    -0.44753  &    1.68321  &    -1.86689  &    --  \\\hline
Pyrochlore, $S=1/2$   &    0.00000  &    0.05334  &    -0.02667  &    0.05507  &    -0.03644  &    0.03260  &    -0.01600$^*$  &    --  \\
Ruby, $S=1/2$   &    0.00000  &    0.05334  &    -0.02667  &    0.05461  &    -0.03767  &    0.02707  &    -0.01267  &    --  \\
Bethe ($n=6$), $S=1/2$   &    0.00000  &    0.05334  &    -0.02667  &    0.05507  &    -0.03656  &    0.03265  &    -0.01683  &    --  \\\hline
Pyrochlore, $S=1$   &    0.00000  &    0.21189  &    -0.10594  &    0.54542  &    -0.47087$^*$  &    --  &    --  &    --  \\
Ruby, $S=1$   &    0.00000  &    0.21189  &    -0.10594  &    0.54611  &    -0.47811$^*$  &    --  &    --  &    --  \\ 
Bethe ($n=6$), $S=1$   &    0.00000  &    0.21189  &    -0.10594  &    0.54541  &    -0.47054  &    --  &    --  &    --  \\\hline
Bethe ($n=8$), $S=1/2$   &    0.00000  &    0.05747  &    -0.02874  &    0.05479  &    -0.03510  &    --  &    --  &    --  \\\hline 
Bethe ($n=8$), $S=1$   &    0.00000  &    0.22650  &    -0.11325  &    0.54815  &    --  &    --  &    --  &    --  \\\hline\hline 
\end{tabular}
\caption{Coefficients of the expansion in powers of $\alpha$ for different clusters and spin. The upper (lower) table lists coefficients of the energy of strong (weak) bonds $E_{\text{strong}}$ ($E_{\text{weak}}$) located on a single plaquette (between two plaquettes). The total energy per site is then given by $E_\alpha = \sum_k E_{\text{strong}}(k) (-\alpha)^k + 2\sum_k E_{\text{weak}}(k) (-\alpha)^k $. 
% Note that each order enters with $\alpha^k$ yielding convergence for small $\alpha$. 
For some of the models, the largest order of the expansion of $E_{\rm weak}$ in $\alpha$ is incomplete and we mark the corresponding coefficient by an asterix $^*$. However, these coefficients only miss a small number of contributions from some of the largest clusters, which we expect to be negligible.
}
\label{tab:coefficients}
\end{table*}

\FloatBarrier

\section{Numerical Linked Cluster Expansion}\label{app:NLCE}

The numerical linked cluster expansion (NLCE) \cite{tang_nlce_2013,schafer_magnetic_2022} was originally developed to compute extensive thermodynamic observables at \emph{finite} temperature in the thermodynamic limit.
Starting from a pre-defined unit, \eg  a hexagon or tetrahedron, the lattice is built up systematically by expanding the clusters unit by unit.
Each generated cluster $c$ has a \emph{weighted} contribution $W\left[\thermal{O_c}\right]$ to the observable $\thermal{O}$ at some inverse temperature $\beta$.
The weight is obtained by subtracting the weights of smaller connected clusters $c^\prime$ that are contained in the cluster $c^\prime\subset c$.
Each cluster has $L_c$ topologically equivalent instances arising during the expansion.
\begin{align}
    &\thermal{O} = \sum_c L_c W\left[\thermal{O_c}\right]\quad\text{with}\quad W\left[\thermal{O_c}\right] = \thermal{O_c} - \sum_{c^\prime \subset c} W\left[\thermal{O_{c^\prime}}\right]\label{eq:NLCE}
\end{align}
The numerical determination of the observable for a single cluster does not allow for approximate methods such as quantum typicality \cite{schnack_magnetism_2018} or finite-temperature DMRG \cite{schafer_pyrochlore_2020} as it is extremely sensitive to errors.
Hence, as far as finite-temperature properties are concerned, observables have to be evaluated using full ED. 
In this letter, however, we also use the NLCE to compute \emph{zero temperature} properties. Since at zero temperature, only the ground state contributes, we can use the Lanczos algorithm (which is numerically exact for $T=0$) to compute the expansion to much larger orders (six instead of three) than would be possible for an expansion at finite temperature.

\begin{figure}
	\includegraphics[width=0.66\columnwidth]{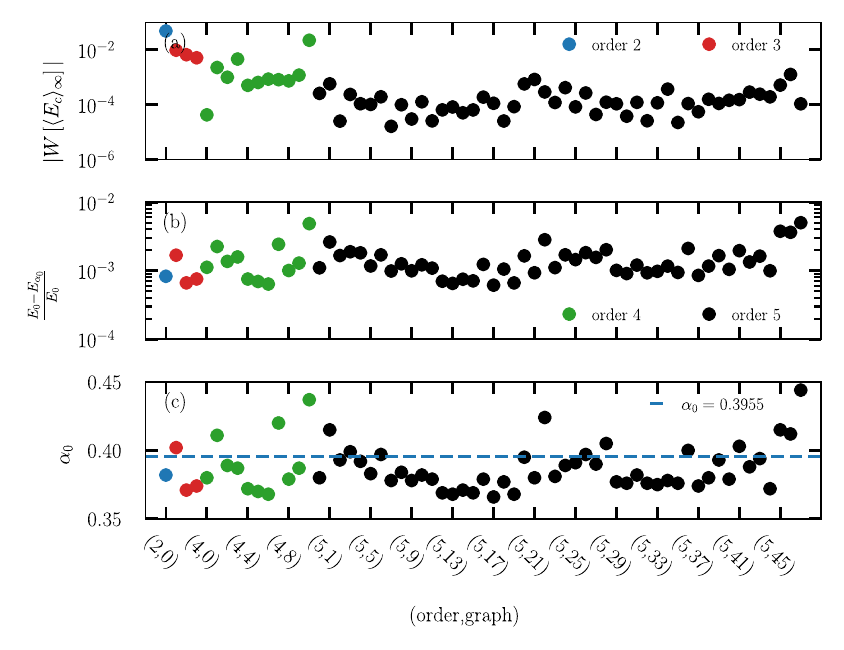}
	\caption{Properties of finite clusters generated within NLCE based on hexagons.  (a) The absolute value of weights for each cluster enters the calculation of the energy at $T=0$, demonstrating the rapid convergence of the expansion. (b) The difference between the exact ground-state energy of each cluster and the variational energy obtained by optimizing $\ket{\Psi_{\alpha_0}} = e^{-\alpha_0V}\ket{\Psi_0}$  on each cluster. The two agree very closely, suggesting that the dressed hexagon state is stable on every individual cluster contributing to the NLCE expansion. (c) The optimal variational parameter, $\alpha_0$ for each cluster, which varies very little between clusters, always agrees closely with the value we obtain in the thermodynamic limit [\autoref{app:time_evol}]. The $x$-axis labels different clusters organized by (order,label).}%%
	\label{fig:finite_clusters}
\end{figure}

The NLCE algorithm is in a spirit similar to a high-temperature expansion, where larger clusters contribute at higher orders in $\beta$. Physically, this is motivated by a growing correlation length at lower temperatures.
Once the correlation length is larger than the largest cluster included in the expansion, consecutive orders will usually diverge away from each other immediately, signaling the failure of convergence. This is seen in Fig. 2 (a) of the main text for the expansion based on tetrahedra (blue), which has proven extremely powerful as it converges down to temperatures $T\approx 0.2J$, inducing a relevant upper bound on the ground-state energy. 
The behavior is quite different if the expansion is instead based on hexagons, shown as orange curves in the same figure. 
While convergence is lost around $T\approx 1J$, the individual curves do not diverge to infinity and the algorithm seems to converge \emph{again} for $T\rightarrow 0$.
Remarkably, the convergence at $T=0$ is reached at the second order in the NLCE expansion.
Hence, the weight of larger clusters and their contribution is very small.
This can be observed in \autoref{fig:finite_clusters} (a), which shows the absolute value of the weight of the energy at $T=0$, $\vert W\left[\langle{E_c}\rangle_{\infty}\right]\vert$, for each cluster from order two to five.
The fact that the individual orders do not diverge to infinity at intermediate temperature suggests that the low energy spectrum for the clusters is similar. 
Other literature results \cite{harris_ordering_1991,hagymasi_possible_2021,astrakhantsev_broken-symmety_2021} and DMRG calculations, \cf \autoref{app:mps}, for the checkerboard lattice as well as the introduced wavefunction propose a ground-state energy that is slightly higher than the result from the NLCE calculation.

As stated before, the loss of convergence is associated with the presence of correlations beyond the largest clusters considered. In light of this, the fact that the series seems to converge at second order in the limit $T\to0$ suggests a simple structure of the ground states of considered clusters.
In particular, in the second order, one considers exactly two coupled hexagons. This, when put on the pyrochlore lattice given a hard-hexagon covering, can capture only a uniform energy density on all hexagons as well as a (different) uniform energy density on the bonds connecting any two hexagons. 
This is fully consistent with the weakly dressed valence-bond crystals proposed in this letter. Note that even though the convergence is basically achieved at order two, we do compute the expansion up to order \emph{six}. 
Higher orders include clusters which in principle \emph{could} break the symmetries of the hard-hexagon states, but computing their ground states shows that they do not. Instead, the ground states of each individual cluster realize a hard-hexagon crystal, described by the variational wave function introduced in \autoref{app:time_evol} to remarkable accuracy, as shown in \autoref{fig:finite_clusters} for the clusters up to order five. 
For each cluster, we optimize the hard-hexagon wavefunction and find the optimal value of the parameter $\alpha$ using imaginary-time evolution. In each case, the resulting energy agrees closely with the exact ground-state energy of the cluster. The optimal values of $\alpha$ hardly vary from cluster to cluster and agree closely with the value of $\alpha_0$ obtained in the thermodynamic limit.
We do not include the $283$ clusters appearing at order six in the figure for clarity. The final NLCE results for the energy at zero temperature at orders two and six are $\text{NLCE}_2=-0.4917J$ and $\text{NLCE}_6=-0.4919J$ indicating the little influence of larger clusters. 
{The \textit{convergence} in the second order is further reflected by the small error in Tab. I of the main text for the different models considered in this work.
It is obtained by the difference between the second and the third order.}

\section{Matrix-Product State Approaches}\label{app:mps}

Approaches based on matrix-product states (MPS) \cite{Verstraete_matrix_2004,schollwock_the_2005,Feiguin_finite_2005,schollwock_the_2011} are the method of choice for strongly correlated one-dimensional systems.
Being a one-dimensional technique, generalizations to higher dimensions are obtained by linearizing the system using a one-dimensional path -- a ``snake'' -- that traverses the system.
The class of models proposed here is well suited for the latter flavor by choosing the snake path to traverse the plaquettes covering the lattice.
We implement such a ``snake'' MPS using the ITensor \cite{fishman_itensor_2022} package.
First, we use it to perform imaginary-time evolution on large finite clusters of the checkerboard and ruby lattices and use this to corroborate the correctness of our determination of the variational energy of the dressed hard-hexagon state (see \autoref{app:time_evol}). Second, we also use DMRG on finite clusters of the checkerboard and ruby lattices to study the structure of their local excitations. The result for the ruby lattice, which is equivalent to the pyrochlore up to the second order in perturbation theory and at larger orders has more closed loops, also yields a heuristic lower bound on the gap of the pyrochlore lattice.

\subsection{Imaginary-Time Evolution}

We use the global subspace expansion for the time-dependent variational principle (TDVP) \cite{mingru_time-dependent_2020} to perform imaginary-time evolution starting from the single-plaquette ground state $\ket{\Psi_0}$.
The main result is shown in \autoref{fig:tdvp_checkerboard}, which compares the variational energy $E_\alpha$, obtained from imaginary-time evolution on the checkerboard lattice using (i) exact diagonalization on a cluster of $N=36$ sites with periodic boundary conditions, (ii) the linked cluster theorem expansion described in \autoref{app:time_evol}, and (iii) TDVP on a finite cluster of $N=100$ sites with periodic boundary conditions.
For the ED result, the dependence of the variational energy on $\alpha$ agrees qualitatively but not quantitatively with the other two results. We attribute this to pronounced finite-size effects. In contrast, the result from linked cluster expansion and from TDVP are in excellent \emph{quantitative} agreement. This confirms the validity of the expansion, which we also use for the pyrochlore lattice where full TDVP on large clusters in unattainable.

\subsection{DMRG results}

\setlength{\tabcolsep}{12pt}
\begin{table}[t]
\centering
\begin{tabular}{l | c  c  c | c c c }\hline \hline 
   &  \multicolumn{3}{c|}{Checkerboard}   & \multicolumn{3}{c}{Ruby}    \\  [1ex] \hline   
    L & $4$ & $5$ & $6$  & $3$ & $4$ & $5$  \\ \hline    
$E_0/N$, $S=1/2$ & -0.51493 & -0.51400 & -0.51364 & -0.49092 & -0.48834 & -0.48669\\ 
$\Delta_{m_z=0}$, $S=1/2$ & -0.44466 & -0.48720 & -0.47179  & -0.24330 & -0.25418 & -0.27083  \\ 
$\Delta_{m_z=\pm 1}$, $S=1/2$ & -0.64910 & -0.62511 & -0.63646  & -0.34568 & -0.33059 & -0.35501 \\\hline  
$E_0/N$, $S=1$ & -1.53672 & -1.53422 & -1.53262  & -1.48757 & -1.48191 & -1.47693  \\ 
$\Delta_{m_z=0}$, $S=1$ & -0.54704 & -0.59843 & -0.68824  & -0.37589 & -0.30215 & -0.36060 \\ 
$\Delta_{m_z=\pm 1}$, $S=1$ & -0.62975 & -0.61802 & -0.55928 & -0.38140 & -0.39578 & -0.43181\\\hline \hline 
\end{tabular}
\caption{DMRG data for checkerboard and ruby lattice ($J=1$). The number of plaquettes is $L^2$ such that the clusters contain $N=4L^2$ and $N=6L^2$ spin-$1/2$ or spin-$1$ sites with periodic boundary conditions, respectively. The one-dimensional ``snake'' path is placed within the plaquettes and breaks the inversion symmetry by choosing one over the other possible tiling in the checkerboard lattice. All results are obtained for a bond dimension of $\chi=4000$, and the maximal truncated weight is of the order $10^{-5}$. $\Delta_{m_z}$ refers to the gap obtained in the magnetization sectors $m_z=0,\pm 1$. The derived variational energy obtained from the cluster expansion is $E_{\alpha_0}=-0.51344$ for the checkerboard and $E_{\alpha_0}=-0.48947$ for the ruby lattice. Note that the variational energy for the ruby lattice is slightly lower than the DMRG calculation.}
\label{tab:gs_energies}
\end{table}

In addition to the imaginary-time evolution, we also utilize DMRG. We compute estimates of the energy and the real-space correlations of the ground state and the first excited state in the $m_z = -1, 0, +1$ magnetization sectors.
We consider finite clusters of both the checkerboard and the ruby lattice, with the linear system size $L$ defined such that the total number of sites in a cluster is $4L^2$ and $6L^2$, respectively. In all cases, we construct the snake path such that it passes through the plaquettes and adjacent plaquettes in real space is traversed by it with a maximal plaquette distance of $2L$ in the one-dimensional topology.
A naive linearization of a cluster with periodic boundaries would yield a maximal distance of $L^2$.

\begin{figure}[t]
     \centering
     \begin{subfigure}[b]{0.49\textwidth}
         \centering
         \includegraphics[width=\textwidth]{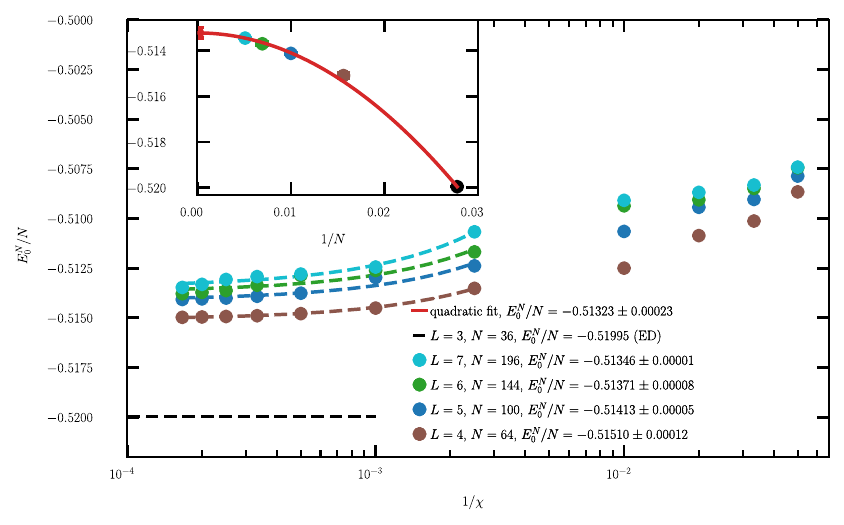}
         \centering
         \caption{Checkerboard, $S=1/2$.}
         \label{fig:gs_energy_checkerboardS1/2}
     \end{subfigure}
        \centering
     \begin{subfigure}[b]{0.49\textwidth}
         \centering
         \includegraphics[width=\textwidth]{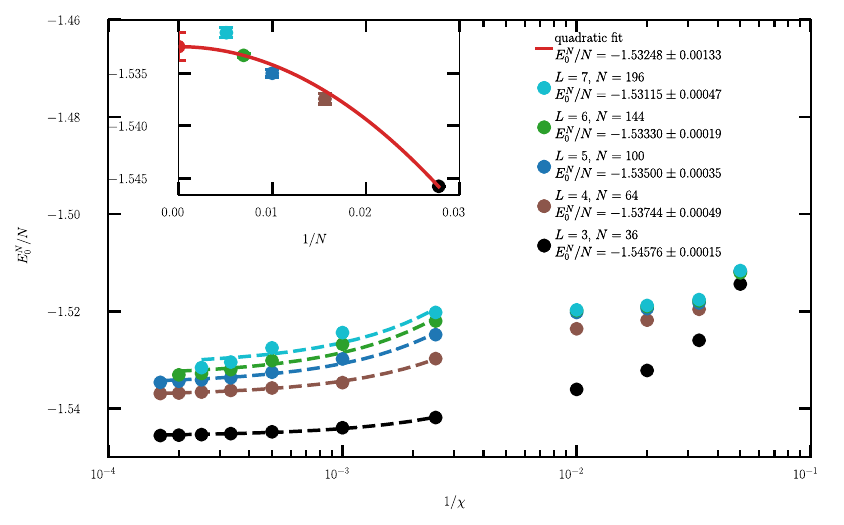}
         \centering
         \caption{Checkerboard, $S=1$.}
         \label{fig:gs_energy_checkerboardS1}
     \end{subfigure}
     \centering
     \begin{subfigure}[b]{0.49\textwidth}
         \centering
         \includegraphics[width=\textwidth]{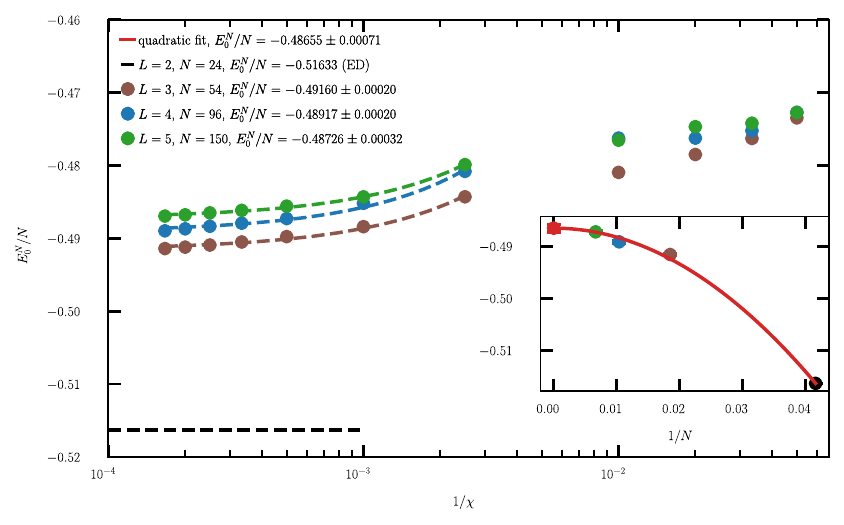}
         \centering
         \caption{Ruby lattice, $S=1/2$.}
         \label{fig:gs_energy_rubyS12}
     \end{subfigure}
        \centering
     \begin{subfigure}[b]{0.49\textwidth}
         \centering
         \includegraphics[width=\textwidth]{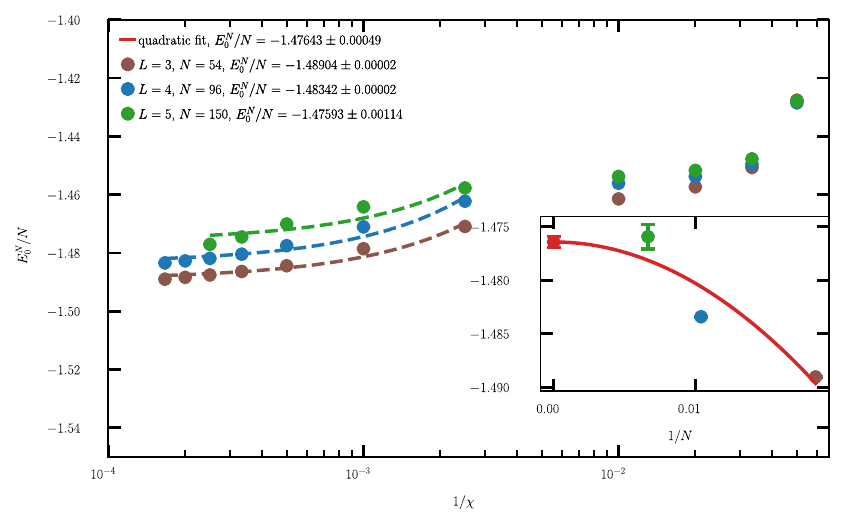}
         \centering
         \caption{Ruby lattice, $S=1$.}
         \label{fig:gs_energy_rubyS1}
     \end{subfigure}
     %\hfill
        \caption{{Extrapolation of the ground-state energy for the checkerboard and ruby lattice for $S=1/2$ and $S=1$ using DMRG data at finite bond dimension. We first extrapolate to infinite bond dimension $1/\chi \rightarrow 0$ and then to the thermodynamic limit $1/N\rightarrow 0$. We use a linear fit in $\frac{1}{\chi}$ to extrapolate to infinite the bond dimension and a quadratic fit for the thermodynamic limit $\frac{1}{N^2}$ to be consistent with the pyrochlore results from Ref.~\cite{hagymasi_possible_2021}.}}        \label{fig:gs_energy_extra}
\end{figure}

Results are shown in \autoref{tab:gs_energies}. The ground-state energy per site is denoted by $E_0/N$.
{We carefully examine the ground-state energy following a similar procedure as carried out in Ref.~\cite{hagymasi_possible_2021}, which serves as an estimate for the pyrochlore $S=1/2$ case. We first extrapolate to infinite bond dimension and then to the thermodynamic limit as illustrated in \autoref{fig:gs_energy_extra}.
To obtain an energy estimate for a cluster of finite size $N$, we extrapolate the energy in $\frac{1}{\chi}$ where is $\chi$ is the bond dimension. 
The error is defined by the difference between the last data point (largest bond dimension) and the extrapolated limit $E_0^N$.
The energy estimates obtained for finite-size clusters are shown in the insets of \autoref{fig:gs_energy_extra} on a $\frac{1}{N}$ scale.
As done in Ref.~\cite{hagymasi_possible_2021}, we use a quadratic fit in $\frac{1}{N^2}$ to determine the energy in the thermodynamic limit that takes the uncertainty of the finite clusters into account.
The confidence interval is obtained by the difference between the largest system size and the extrapolated result $E_0$.
}
We find the ground-state estimate is in excellent agreement for all lattices and spin lengths with the variational estimates $E_\alpha$ (\cf in Tab. 1 of the main text).
This applies not only for the ground state but also for the reals-space energy \emph{distribution} on the lattice, that is, the energy densities on strong bonds ($E_{\rm strong} = \expval{H_p}_{\alpha_0}/L_p$) and weak bonds ($E_{\rm weak} = \expval{V_e}_{\alpha_0}/4$).
The individual contribution of weak and strong bonds to the energy are summarized in \autoref{tab:energies}.

\begin{figure}[t]
     \centering
     \begin{subfigure}[b]{0.49\textwidth}
         \centering
         \includegraphics[width=\textwidth]{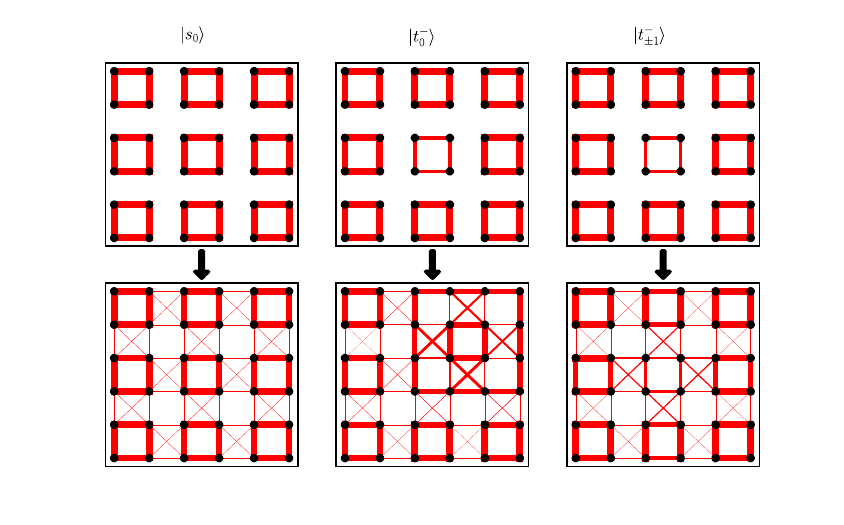}
         \centering
         \caption{Checkerboard, $S=1/2$.}
         \label{fig:real_space_checkerboard}
     \end{subfigure}
     %\hfill
     \begin{subfigure}[b]{0.49\textwidth}
         \centering
         \includegraphics[width=\textwidth]{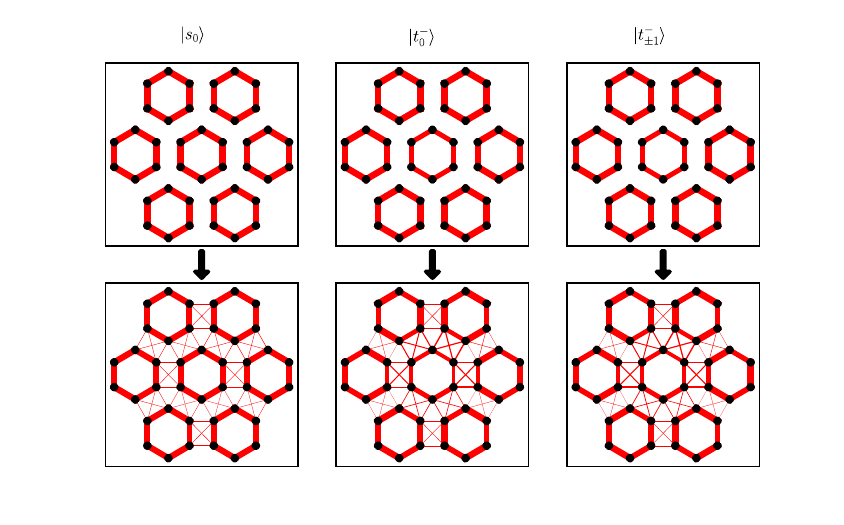}
         \caption{Ruby lattice, $S=1/2$.}
         \label{fig:real_space_ruby}
     \end{subfigure}
        \caption{Real-space correlation obtained by DMRG for the (a) checkerboard ($L=5$, $N=100$) and (b) ruby ($L=4$, $N=96$) lattice. {{The plots show the truncated lattice around the localized excitation.}} The thickness of the red lines is proportional to the energy contribution of each individual bond. The top rows show the initially prepared product states restricted to the plaquettes. The first column represents the ground states of $H_0$ (top) and $H_0+V$ (button), and the second and third column exhibit a triplet excitation, $\ket{t^-_0}$ and $\ket{t^-_{\pm 1}}$, in the $m_z=0$ and $m_z=\pm 1$ sector initially. The lower rows show the ground state and the respective excitations in the different magnetization sectors of the full system. The data were obtained for a bond dimension $\chi=4000$ using $U(1)$ DMRG, and the truncation error is of order $10^{-5}$. As DMRG does not conserve the total spin and the $m_z=0$ gap differs from the $m_z=\pm 1$ gap, the observed $m_z=0$ excitation is a singlet state.}        \label{fig:real_space}
\end{figure}

Motivated by these excellent results for the ground state, we now turn to a careful study of the lowest-lying excitations.
We initially prepare the lowest lying excitation on a single plaquette which is given by the triplet $\ket{t^-_{m_z}}$ for the respective magnetization sectors $m_z=0,\pm 1$ on top of the of a product state of single plaquette ground states $\ket{\Psi_0}$.
The resulting triplet gap is $1J$ for an isolated square and $\sim 0.69J$ for an isolated hexagon.
DMRG preserves the total magnetization such that the lowest lying excitation in the $m_z=1$ sector is obtained by minimizing the energy without constraints.
In contrast, the $m_z=0$ excitation has to be orthogonalized to the plaquette ground state.
For both models, we found a lower gap in the $m_z=0$ sector compared to $m_z=\pm 1$. This suggests that the lowest lying excitation in both cases is a \emph{singlet}. This is somewhat surprising, given that the lowest lying excitation on both, a single square as well as a single hexagon, is a triplet.
Possible scenarios include breaking plaquettes into different (larger) unfrustrated motifs or strongly dressed localized singlets excitation.
\autoref{fig:real_space} shows the real space correlation of the individual states for the checkerboard lattice with $L=5$ and the ruby lattice with $L=4$.
The left columns show the respective ground states followed by the lowest-lying excitations in the $m_z=0$ and $m_z=\pm 1$ sectors.
In the case of the checkerboard lattice, the lowest lying excitation in the $m_z=0$ sector indeed forms a larger motif [bottom center of \autoref{fig:real_space_checkerboard}]. 
In contrast, the $m_z=1$ excitation appears localized on a single plaquette for both considered models, consistent with a weakly dressed, localized triplet excitation. 
The initial triplet gaps ($1J$ and $\sim 0.69J$) are reduced significantly by an increasing contribution of the neighboring weak bonds.

Our results on the ruby lattice suggest a triplet gap of size $\Delta_{m_z = \pm1} \sim 0.36J$. We note that this is similar in value to estimates on the pyrochlore lattice using DMRG calculations on a $N=64$ cluster, yielding $\sim 0.42J$ \cite{hagymasi_magnetization_2022} and variational Monte Carlo yielding $\sim 0.40J$ \cite{astrakhantsev_broken-symmety_2021}.

\setlength{\tabcolsep}{6pt}
\begin{table*}[t]
\centering
\begin{tabular}{c  | c c c | c c c | c c c }\hline\hline  
   & \multicolumn{3}{c|}{Pyrochlore }   & \multicolumn{3}{c|}{Ruby}   & \multicolumn{3}{c}{Checkerboard}   \\ [1ex] \hline   
    & $E_{\text{strong}}$ & $E_{\text{weak}}$ & $E_0$     & $E_{\text{strong}}$ & $E_{\text{weak}}$ & $E_0$  & $E_{\text{strong}}$ & $E_{\text{weak}}$ & $E_0$ \\ [1ex] 
\hline
$e^{-\alpha_0 V}\vert 0\rangle $ & -0.4301 & -0.0297 & -0.4895  & -0.4302 & -0.0297 & -0.4894 & -0.4801 & -0.0167 & -0.5134 \\
DMRG & -- & -- & -0.4898 & -0.4357 & -0.0264 & -0.4885 & -0.4758 &  -0.0190 &  -0.5139 \\
NLCE$_1$ & -0.4671 & -- & -0.4671 & -0.4671 & -- & -0.4671 & -0.5 & -- & -0.5\\
NLCE$_2$ & -0.4235 & -0.0341 & -0.4917& -0.4235 & -0.0341 & -0.4917 & -0.4792 & -0.0173 & -0.5139 \\ 
ED OBC  & -0.4272 & -0.0337 & -0.4945 & -0.4337 & -0.0267 & -0.4871 & -0.4798 & -0.0176 &-0.5150\\ 
\hline\hline
\end{tabular}
\caption{Contribution of strong and weak bonds to the energy ($S=1/2$, $J=1$) of the hard-hexagon state, obtained by different methods. The first row refers to the imaginary-time evolution described in \autoref{app:time_evol} and the second row presents DMRG results which are obtained within this work, \cf \autoref{app:mps}, for the ruby and checkerboard lattice (bond dimension $\chi=4000$ and truncated error is of order $10^{-6}$) and by Ref. \onlinecite{hagymasi_possible_2021} for the pyrochlore lattice. Row three and four refer to the first and second-order NLCE expansion. The last row presents ED results for a finite system with open boundary conditions ($N=42$ and $N=20$) where a center plaquette of length $n$ is coupled to $n$ other plaquettes ($N=(n+1)n$) via the doubly frustrated quartets. The outer plaquettes exhibit open boundaries and the strong and weak bonds are extracted from the center plaquette. This reflects the robustness of the plaquette crystal.}
\label{tab:energies}
\end{table*}

\section{Hard-Hexagon Coverings of the Pyrochlore Lattice}

\subsection{Finding Hexagon Coverings by Simulated Annealing}

A hard-hexagon covering of the pyrochlore lattice is a choice of non-overlapping hexagons such that each vertex is part of exactly one hexagon.
In all such coverings, one example is shown in Fig. 1 (a) (main text), the hexagons are arranged within one of the $\{001\}$ planes which are then stacked along the third direction. Each plane can be translated independently by either one of the $\langle110\rangle$ or $\langle1\bar10\rangle$ directions while preserving the hard-covering constraint. This yields an exponential, but subextensive number of coverings $N_{\rm cover}= 3 \times 2^{4L/3}$ where $L$ is the linear system size in units of cubic, 16-site, unit cells.
As the main text explains, these coverings break rotation, translation, and inversion symmetry. The translation symmetry is particularly striking as its unusual periodicity (three) means that most finite clusters used in numerical studies of the pyrochlore Heisenberg model to date are \emph{incommensurate} with any hard-hexagon covering.

We have verified that the planar coverings as discussed above and in the main text are the only possible coverings by classical Monte-Carlo simulation of a hard-hexagon model
\begin{equation}
    H_{\rm hex} = \sum_j (n_{\rm hex}(j) - 1)^2,
\end{equation}
where the sum is over sites $j$ and $n_{\rm hex}(j)$ is the number of occupied hexagons that the site is part of. The zero-energy states of this model are exactly the hard-hexagon coverings of the pyrochlore lattice. We performed $10^4$ simulated annealing runs on a system with $6\times6\times6$ unit cells (that is $N_{\rm v} = 3456$ sites) and found all 68 non-symmetry-equivalent hexagon coverings of the expected form and nothing else (note that $10^3$ runs already suffice to find all non-symmetry equivalent coverings). We take this as strong evidence that indeed all possible coverings of the pyrochlore lattice with hard hexagons take the form shown in Fig. 1 (a).
For $L=12$ (that is $N_{\rm v} = 27648$ sites), we are not able to find \emph{all} possible symmetry-equivalent planar states due to their exponentially large number.
However, we did perform $10^4$ simulated annealing runs and verified that all coverings found are of the expected, planar form. 

\subsection{Dimer Structure Factor}

One way do diagnose such a hard-hexagon valence-bond-crystal state would be the unusual translational symmetry breaking with period three. Since two-spin correlators vanish beyond single hexagons, one needs higher-weight correlation functions. A simple, transitional invariant quantity to diagnose such a symmetry breaking would be the dimer-dimer structure factor
\begin{align}
    S_{\rm dimer}(\vec q) =& \sum_{\expval{ij}, \expval{kl}} \exp(-i \vec q \cdot \left[ \frac{1}{2}(\vec r_i + \vec r_j) - \frac{1}{2}(\vec r_k + \vec r_l) \right])
    ~\expval{\vec S_i \cdot \vec S_j\, \vec S_k \cdot \vec S_l},
    \label{eq:dimer_s_q}
\end{align}
where $\expval{ij}$ and $\expval{kl}$ are nearest-neighbor pairs on the lattice. We show this quantity in Fig. 3 of the main text. 
To compute it in practice, we approximate the four-spin correlation function in \autoref{eq:dimer_s_q} as
\begin{equation}
    \expval{\vec S_i \cdot \vec S_j\, \vec S_k \cdot \vec S_l} =
    \begin{cases}
        \expval{\vec S_i \cdot \vec S_j\, \vec S_k \cdot \vec S_l} & \text{if $\expval{ij}$, $\expval{kl}$ on same hexagon} \\
        \expval{\vec S_i \cdot \vec S_j}\expval{\vec S_k \cdot \vec S_l} & \text{else} \\
    \end{cases}.
    \label{eq:dimer_s_q_approx}
\end{equation}
That is we include the dominant short-range correlations exactly while approximating the long-range correlations as the Fourier transform of the energy density. This is readily computed for a specific hexagon covering and assuming self-averaging, we then obtain the plots in Fig. 3 of the main text by computing \autoref{eq:dimer_s_q_approx} for all coverings of a $L=6$ ($N_{\rm v}=3456$) pyrochlore cluster with stacking direction $[001]$ and taking the average.

To include longer-ranged correlations is possible, but including for example also neighboring hexagons and bonds connecting them exactly only change $S_{\rm dimer}(\vec q)$ by about $5\%$. In fact, the Bragg peaks and even the high-intensity broad features in Fig. 3 of the main text are already resolved without even treating correlations within one hexagon exactly. This is needed only to resolve the small-intensity broad features between resonances.

\subsection{Structure factor}
\begin{figure}
    \centering
    \includegraphics{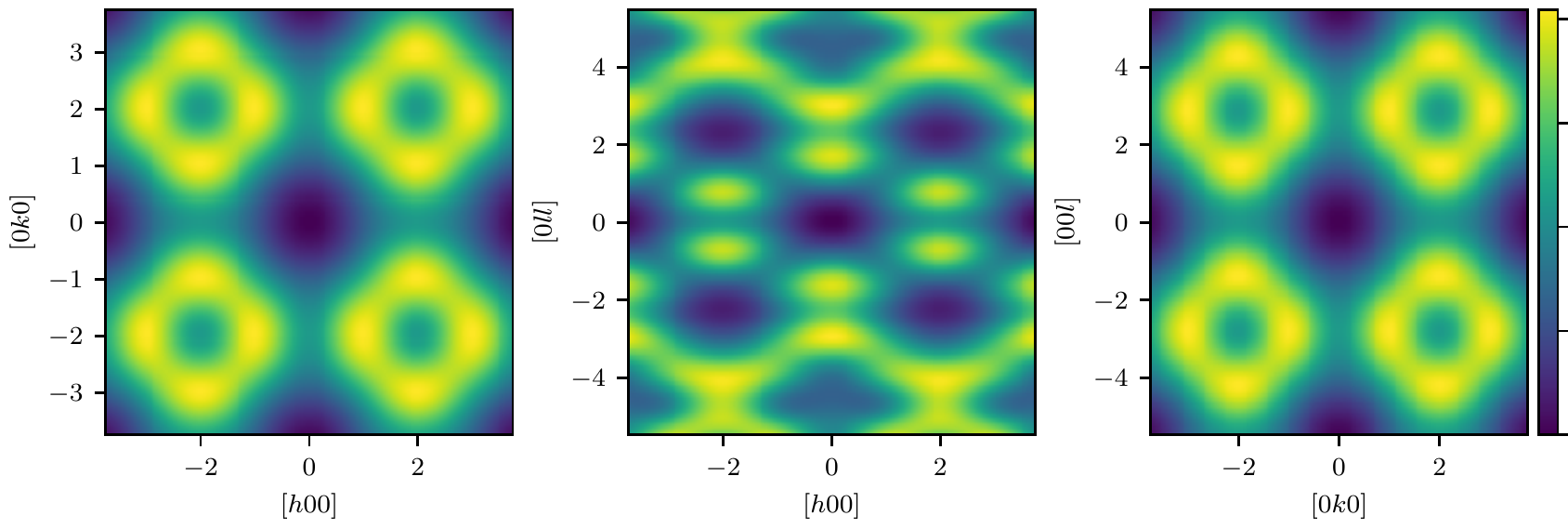}
    \caption{Structure factor $S(\vec q)$ of the undressed hard-hexagon state $\ket{\Psi_0}$, with stacking direction [001]. The difference to Ref. \onlinecite{lee_emergent_2002} comes from the fact that we consider $S=1/2$ rather than classical vector spins and also we also do not multiply by any magnetic form factor.}
    \label{fig:spin_s_q}
\end{figure}

For completeness, we show in \autoref{fig:spin_s_q} the spin-spin structure factor
\begin{equation}
    S(q) = \sum_{i,j} \exp[-\vec q \cdot (\vec r_i - \vec r_j)] \expval{\vec S_i\cdot \vec S_j}
\end{equation}
of the undressed hard-hexagon state $\ket{\Psi_0}$. In the undressed case, all inter-hexagon correlation functions vanish and we only have to average the structure factor of a single hexagon over the four possible orientations that it can have in the pyrochlore lattice.

\section{Kinetic Energy from Multiboson Theory}\label{app:boson}

In the following, we describe the multiboson theory approach \cite{sachdev1990, romhanyi2012} used to compute the kinetic energy of triplet excitations up to first order in $V$, as shown in Fig. 4 (a) of the main text. 

We start from a general Hamiltonian as decomposed into a term $H_0$ constituting the individual hexagons and a term $V$, which couples these hexagons via tetrahedra. To fix notation, we write explicitly
\begin{subequations}
\begin{align}
    H &= H_0 + V, \\
    H_0 &= \sum_{h \in \{\rm hexagons\}} \sum_{\expval{ij}\in h} \vec S_{h, i} \cdot \vec S_{h, j}, \\
    V &= \sum_{t=\expval{h, h'} \in \{\rm tetrahedra\}} \sum_{\expval{ij}\in t} \vec S_{h, i} \cdot \vec S_{h', j}.
\end{align}
\label{eq:hamiltonian-multiboson0}
\end{subequations}
where we have chosen to split up the site index $r$ into the index of the hexagon $h$ the site belongs to and a site index within that hexagon $j \in \{1, \dots 6\}$, that is $r := (h, j)$. The notation $t=\expval{h, h'}$ implies that the tetrahedron $t$ couples hexagons $h$ and $h'$.

We now proceed similarly to the bond-operator formalism of Ref. \onlinecite{sachdev1990} to rewrite the spin operator in the basis of single-hexagon eigenstates. We denote the state $n\in\{0, \dots, 63\}$ on hexagon $h$ as $\ket{h, n}$ and define a set of bosons $a_{h, n}$ such that
\begin{align}
    \ket{h, n} &= a^\dagger_{h, n} \ket{\rm vacuum}, \\
    S^\alpha_{h, j} &= \left( S^\alpha_{h, j} \right)_{nm} \ketbra{h, n}{h, m}, \\
        & = \left( S^\alpha_{j} \right)_{nm} a_{h, n}^\dagger a_{h, m},
\end{align}
where
\begin{equation}
    \big( S^\alpha_{j} \big)_{nm} = \bra{n} S^\alpha_{j} \ket{m} = \bra{h, n} S^\alpha_{h, j} \ket{h, m}\, \forall h
\end{equation}
and $\alpha=x,y,z$. We have used that this matrix element does not depend on the hexagon $h$ but only on the site index $i,j$.
Then, while $H_0$ is quadratic in the $a_n$, the coupling $V$ consists of four-boson terms.

The bosons are constrained such that there must be exactly one boson on each hexagon
\begin{equation}\label{eq:hardcore-bosons}
    \sum_m \hat n_{h, m} = 1 \quad
    \forall \ h
\end{equation}

The idea of the linear multiboson theory is that, similar in spirit to linear spin wave theory, one can derive a quadratic approximation to $H$, assuming the excitation density is small. That is to say, we assume
\begin{equation}
    a_{h, m}^\dagger a_{h, m} = \hat n_{h, m} \ll 1 \text{ for } m > 0.   
\end{equation}
\autoref{eq:hardcore-bosons}
then implies that $\hat n_{h, 0} \approx 1$. 
The desired quadratic approximation is then obtained by neglecting all terms that are not at least quadratic in the ground state Boson operators and setting all quadratic terms to be $1$. 
Formally, this can be done by substituting the ground state Boson operators with a number
\begin{align}
    a_{h, 0} &\to \sqrt M,\\
    a_{h, 0}^\dagger &\to \sqrt M,
\end{align}
and keeping only of at least order $\order{M}$. Letting $M\to1$ then yields the desired quadratic approximation to $H$
\begin{align}
    &H_0 = \sum_h \sum_n \omega_n a_{h, n}^\dagger a_{h, n} \\
  &V \,\,\,= \sum_{t = \expval{h, h'}} \sum_{n, m} \sum_{\expval{ij} \in t} \sum_\alpha  \left( S_{i}^\alpha \right)_{0n} \left( S_{j}^\alpha \right)_{0m} a_{h, n} a_{h',m}+\left( S_{i}^\alpha \right)_{n0} \left( S_{j}^\alpha \right)_{0m} a_{h, n}^\dagger a_{h',m} &&\nonumber \\
  &\phantom{V = \sum_{t = \expval{h, h'}} \sum_{n, m} \sum_{\expval{ij} \in t} \sum_\alpha  }\,\,\,\,\,+\left( S_{i}^\alpha \right)_{0n} \left( S_{j}^\alpha \right)_{m0} a_{h, n} a_{h',m}^\dagger+       \left( S_{i}^\alpha \right)_{n0} \left( S_{j}^\alpha \right)_{0m} a_{h, n}^\dagger a_{h',m}^\dagger &&
\end{align}
where the $\omega_n$ are just the eigenenergies of a single hexagon.

Hence, we have obtained from $H$ a standard (anomalous) bosonic hopping problem that is readily solved by Fourier transform.

\section{Application to other models}
{We briefly discuss the application of the variational wavefunction to other models experiencing the same heuristic of unfrustrated loops with highly frustrated couplings.
In particular, the same physics is observed in the ruby and checkerboard lattice [Fig. 1 in the main text], which have been discussed throughout the supplementary material.
The ground state energies for each case obtained by our variational construction are compared with DMRG and second-order NLCE in Tab. I of the main text.}

{To begin with, we consider the $S=1$ version of the pyrochlore Heisenberg model.
In this case, the entire argument carries over from the $S=1/2$ case, with the additional twist that for integer spins the gap on an even-length loop is not an effect of its finite length~\cite{haldane_nonlinear_1983}.
We obtain a variational energy slightly above a recent estimate from a DMRG study on a cluster of 48 sites \cite{hagymasi_enhanced_2022}, but given the likely considerable finite-size effects on the DMRG result, this constitutes a competitive estimate.}

{For the ruby lattice, there is a unique hard-hexagon covering, preserving all lattice symmetries. The hard-hexagon state on this model is thus a featureless quantum paramagnet, comparable in character to the ground state of the Shastry-Sutherland model \cite{shastry1981}.}

{Finally, for the checkerboard lattice, the shortest unfrustrated loops that tile the lattice are squares. There are two possible hard-square coverings, and thus the hard-square state is a plaquette crystal breaking a $\mathbb{Z}_2$ symmetry. This is consistent with previous results for the ground state order on the checkerboard lattice \cite{sindzingre02, fouet_planar_2003,brenig_planar_2002, chan2011, bishop2012} and so, in this case, our approach confirms the established consensus.}

{In all cases, the  multiboson theory predicts gapped flat bands for the low-lying triplet excitations.
In the case of the checkerboard and ruby lattices, we can use  DMRG\cite{Verstraete_matrix_2004,schollwock_the_2005,Feiguin_finite_2005,schollwock_the_2011}, \cf \autoref{app:mps}, to conduct an additional test of the robustness of the gap~\cite{stoudenmire_studying_2012}.
The knowledge of the (possible) ground state, together with the irrelevance of \emph{kinetic} energy, enables us to use DMRG to reliably estimate the robustness of the gap to the local dressing of the lowest-lying excitations.  
We find that the gaps remain robust, with $\Delta\approx0.47J$ for the checkerboard lattice and $\Delta\approx0.27J$ for the ruby lattice. 
The latter is particularly interesting since it is up to second order in perturbation theory equivalent to the pyrochlore lattice. Even at higher order, the ruby lattice has more closed loops than the pyrochlore, and hence its gap serves as a heuristic lower bound for that of the pyrochlore.}

{Note that this construction is not straightforwardly generalized to arbitrary frustrated models. For example, while the kagom\'e lattice famously exhibits the hexagonal loop motif, it neither allows a hard-hexagon covering nor is the coupling between the hexagons doubly frustrated.}
\bibliography{pyrochlore}